\documentclass[structabstract]{aa}  
\usepackage{graphicx}
\usepackage{txfonts}
\usepackage{natbib}
\bibpunct{(}{)}{;}{a}{}{,} 

\def\lsxf{$L_{\rm{soft X}}/L_{\rm{FIR}}$}

\def\lxratt{${L_{0.4-2.4\rm{\,keV}}^{HXcomp}/L_{0.4-2.4\rm{\,keV}}^{\rm{tot}}}$}
\def\lfir{$L_{\rm{FIR}}$}
\def\xeff{$\epsilon_{\rm{xeff}}$}

\def\ergs{erg s$^{-1}$}
\def\ergsa{erg s$^{-1}$ \AA$^{-1}$}
\def\ergsacm{erg s$^{-1}$ cm$^{-2}$ \AA$^{-1}$}
\def\ergscm{erg s$^{-1}$ cm$^{-2}$}
\def\msun{M$_\odot$}
\def\ebv{{\em E(B-V)}}
\def\ebvneb{\ebv$_{\rm{neb}}$}

\def\lyalp{{Ly$\alpha$}}

\def\luv{{$L_{\rm{UV}}$}}

\def\lcuva{{$L_{1500}$}}

\def\ewhb{{$EW(\rm{H}\beta)$}}

\def\hal{{$\rm{H}\alpha$}}
\def\hb{{$\rm{H}\beta$}}

\def\lha{{$L(\rm{H}\alpha)$}}

\def\oxiil{O~[II]$3727$}
\def\oxiiid{O~[III]$4959$, $5007$}
\def\siivd{{Si~IV $\lambda\lambda1394$,~$1403$ \AA}}
\def\civd{{C~IV $\lambda\lambda1548$,~$1551$ \AA}}

\def\lhalx{\lha$/L_{0.4-2.4\rm{\,keV}}$}
\def\fuviras{{$F_{1529}$}}
\def\chb{{$C(\rm{H}\beta)$}}
\def\ri{$R_{1}$}
\def\lyas{{Ly$\alpha{\rm{S}}$}}

\newcommand{\fesclya}{$f_{\mathrm{esc}}^{\mathrm{Ly}\alpha}$}

\begin{document}
   \title{Physical properties and evolutionary state of the Lyman alpha emitting starburst
galaxy IRAS~08339+6517}


\titlerunning{\lyalp\ emission and physical properties of IRAS~08339+6517}

   \author{H. Ot\'{\i}-Floranes\inst{1,2,3}
          \and
          J.M. Mas-Hesse\inst{2}
          \and
          E. Jim\'{e}nez-Bail\'{o}n\inst{1}
          \and
          D. Schaerer\inst{4,5}
          \and
          M. Hayes\inst{6,5}
          \and
          G. \"{O}stlin\inst{7}
          \and
          H. Atek\inst{8}
          \and
          D. Kunth\inst{9}
}

   \offprints{J.M. Mas-Hesse}

   \institute{Instituto de Astronom\'{i}a, Universidad Nacional Aut\'{o}noma de
M\'{e}xico, Apdo. Postal 106,
	     Ensenada B. C. 22800, Mexico\\
             \email{otih@astrosen.unam.mx}
\and
	     Centro de Astrobiolog\'{\i}a (CSIC--INTA), Departamento de Astrof\'{\i}sica,
POB 78, 
	     E--28691 Villanueva de la Ca\~nada, Spain\\
             \email{mm@cab.inta-csic.es}
\and
	     Dpto. de F\'{\i}sica Moderna, Facultad de Ciencias, Universidad de Cantabria,
39005 Santander, Spain
\and
	     Observatoire de Gen\`{e}ve, Universit\'{e} de Gen\`{e}ve, 51 Ch. des
Maillettes, 1290 Versoix, Switzerland
\and
	     CNRS, Institut de Recherche en Astrophysique et Plan\'{e}tologie, 14 avenue
Edouard Belin, F-31400 Toulouse, France
\and
	     Universit\'{e} de Toulouse, UPS-OMP, IRAP, Toulouse, France
\and
	     Department of Astronomy, Oskar Klein Centre, Stockholm University, SE - 106
91 Stockholm, Sweden
\and
	     Laboratoire d'Astrophysique, \'{E}cole Polytechnique
F\'{e}d\'{e}rale de Lausanne (EPFL), Observatoire, CH-1290 Sauverny, Switzerland
\and
	     Institut d'Astrophysique de Paris (UMR 7095: CNRS \& UPMC), 98 bis Bd Arago,
75014 Paris, France
             }

   \authorrunning{Ot\'{\i}-Floranes et al.}
   
   \date{Received; accepted}

 
  \abstract
   {Though \lyalp\ emission is one of the most used tracers of massive star formation at
high redshift, it is strongly affected by neutral gas radiation transfer effects. A
correct understanding of these effects is required to properly quantify the star formation
rate along the history of the Universe.} 
   {We aim to parameterize the escape of \lyalp\ photons as a function of the galaxy
properties, in order to properly calibrate the \lyalp\ luminosity as a tracer of star
formation intensity at any age of the Universe.} 
   {We are embarked in a program to study the properties of the \lyalp\ emission
(spectral profile, spatial distribution, relation to Balmer lines intensity,...)
in a number of starburst galaxies in the Local Universe. The study is based on
HST spectroscopic and imaging observations at various wavelengths, X-ray data
and ground-based spectroscopy, complemented with the use of evolutionary
population synthesis models.} 
   {We present here the results obtained for one of those sources,
IRAS~08339+6517, a strong \lyalp\ emitter in the Local Universe which is
undergoing an intense episode of massive star formation. We have characterized
the properties of the starburst, which transformed $1.4\times10^{8}$ \msun\ of
gas into stars around $5-6$ Myr ago. The mechanical energy released by the
central Super Stellar Cluster (SSC) located in the core of the starburst has
created a cavity devoid of gas and dust around it, leaving a clean path through
which the UV continuum of the SSC is observed, with almost no extinction. While
the average extinction affecting the stellar continuum is significantly larger
out of the cavity, with \ebv=0.15 in average, we have not found any evidence for
regions with very large extinctions, which could be hiding some young, massive
stars not contributing to the global UV continuum. The observed soft and hard
X-ray emissions are consistent with this scenario, being originated by the
interstellar medium heated by the release of mechanical energy in the first
case, and by a large number of active High Mass X-ray Binaries (HMXBs) in the
second. In addition to the central compact emission blob, we have identified a
diffuse \lyalp\ emission component smoothly distributed over the whole central
area of IRAS~08339+6517. This diffuse emission is spatially decoupled from the
UV continuum, the \hal\ emission or the \hal/\hb\ ratio. Both locally and
globally, the \lyalp/\hal\ ratio is lower than the Case B predictions, even
after reddening correction, with an overall \lyalp\ escape fraction of only
4\%.}
   {We conclude that in IRAS~08339+6517 the \lyalp\ photons resonantly scattered
by an outflowing shell of neutral gas are being smoothly redistributed over the
whole central area of the galaxy. Their increased probabibility of being
destroyed by dust would explain the low \lyalp\ escape fraction measured. In any
case, in the regions where the diffuse \lyalp\ emission shows the largest
\lyalp/\hal\ ratios, no additional sources of \lyalp\ emission are required,
like ionization by hot plasma as proposed for Haro~2, another galaxy in our
sample. These results stress again the importance of a proper correction of
scattering and transfer effects when using \lyalp\ to derive the star formation
rate in high-redshift galaxies.}


   \maketitle

\begin{figure*}
\centering
\includegraphics[height=9 cm,bb=25 135 475 656
dpi,clip=true]{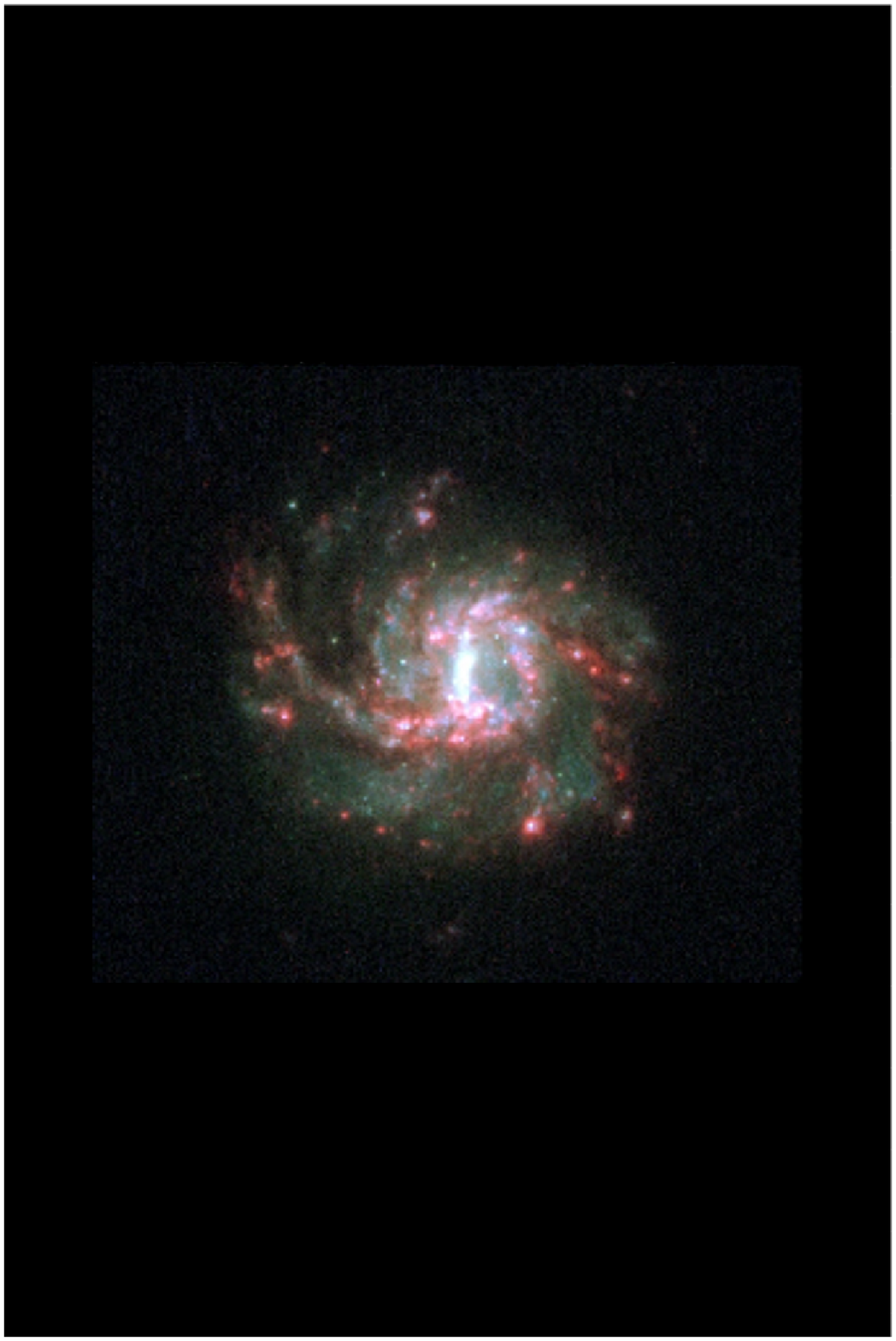}
\includegraphics[height=9 cm,bb=14 14 519 470
dpi,clip=true]{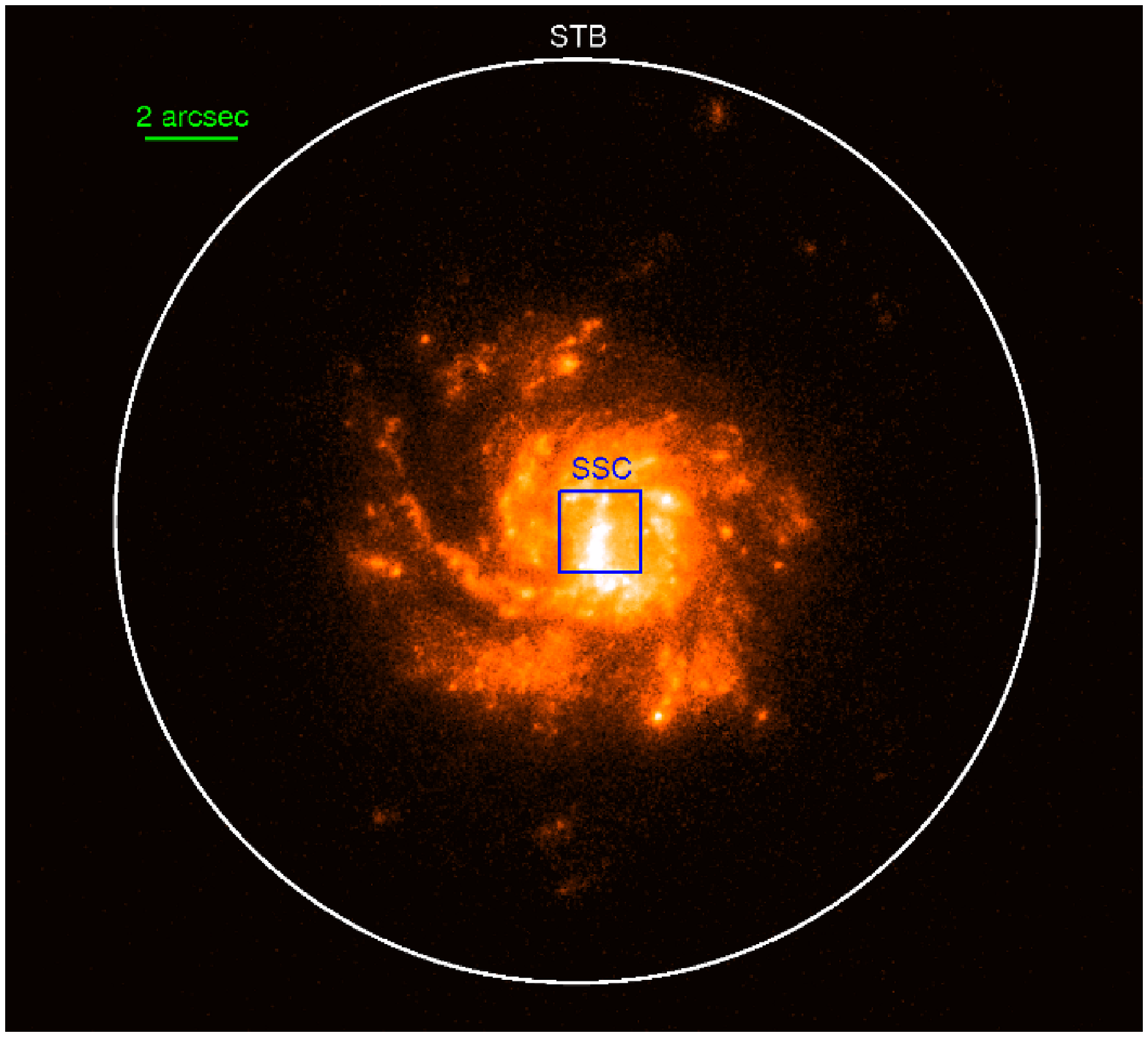}
\caption[Morphology of IRAS~0833]{{\bf Left:} multispectral HST/ACS image of
IRAS~0833. Blue: F300W, green: F435W, red: FR656N (Courtesy of \'Angel R.
L\'opez-S\'anchez -- AAO). {\bf Right:} F140LP UV image. Integrated emission of
the starburst (STB) and central Super Stellar Cluster (SSC) are labelled.
Intensity scale is logarithmic. Same spatial scale in both images, as marked on
the right panel. North is up and east is left.}
\label{figirasuv}
\end{figure*}

\begin{figure}
\centering
\includegraphics[width=9 cm,bb=0 0 754 691
dpi,clip=true]{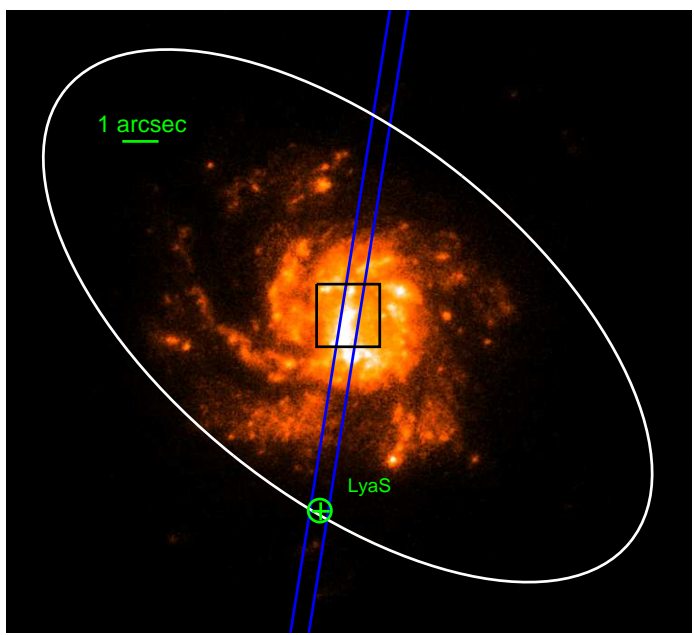}
\caption[GHRS and IUE apertures of IRAS~0833]{F140LP UV image. HST/GHRS
(squared) and IUE (oval) apertures are superimposed, together with the location
of the STIS slit. The width of the apertures and the slit are represented at
real scale. The location of the region \lyas\ discussed in
Sect.~\ref{disclyairas} is marked with a cross within a circle. Intensity scale
is logarithmic. North is up and east is left.}
\label{figirasaper}
\end{figure}

\section{Introduction} 

The \lyalp\ emission is one of the most useful tracers of massive star formation
in the early Universe, since at high redshift it becomes visible in the
optical-infrared range. Moreover, for redshifts above $z\sim4$ it becomes the
only emission line easily acccesible from ground--based telescopes. But the
study of \lyalp\ emission in starburst galaxies in the Local Universe has shown
that radiation transfer through neutral hydrogen can completely distort its
properties, affecting its profile and intensity, and even transforming the
emission into a damped absorption when the column density of neutral gas is high
enough \citep{MasHesse03}. In order to  investigate further the properties of
the \lyalp\ emission, and its correlation with other characteristics of the
starbursts originating it, we are embarked in a program to study in detail
\lyalp\ emitting galaxies in the Local Universe \citep{Ostlin09,Oti12,Hayes13a}.
The final aim of this program is the proper calibration of \lyalp\ as a tracer
of star formation, which could be used to derive the star formation intensity
along the history of the Universe. In this work we present a detailed
multiwavelength analysis of the \lyalp\ emitting galaxy IRAS~08339+6517,
complementing previous partial results by \citet{MasHesse03} and
\citet{Ostlin09}.

IRAS~08339+6517 (hereinafter IRAS~0833) is a face-on spiral galaxy at $80.2$
Mpc, which is undergoing a starburst episode and shows a prominent \lyalp\
emission line. The brightest UV knots are aligned along an internal bar oriented
close to the $N-S$ direction \citep{Ostlin09}. \citet{LopezSanchez06} derived an
age of $4-6$ Myr for the most recent burst using \hal\ spectroscopic data. They
also estimated that there exist two older underlying stellar populations with
ages $100-200$ Myr and $1-2$ Gyr. \citet{Cannon04} reported from NRAO VLA data
that IRAS~0833 harbours $1.1 \times 10^{9}$ \msun\ of neutral hydrogen. They
also detected a tidal stream between IRAS~0833 and its minor companion 2MASX
J08380769+6508579, estimating for the stream  material a mass of $3.8 \times
10^{9}$ \msun. The \lyalp\ emission of IRAS~0833 was first reported by
\citet{Kunth97} from the analysis of HST/GHRS high-resolution spectrum of its
nuclear region. They found that the \lyalp\ line shows a P~Cyg profile, as well
as the presence of a blueshifted secondary \lyalp\ emission component, on top of
the spectral P~Cyg trough. From the same data, \citet{Kunth98} measured the flux
of the line, as well as the column density of the neutral hydrogen which
produces the observed P~Cyg absorption. Later, HST/STIS long-slit observations
with high spectral resolution led \citet{MasHesse03} to confirm that \lyalp\
photons can actually escape from IRAS~0833 because most of the neutral gas
surrounding the source is being pushed out by the starburst activity. The
velocity shift causes that, whereas the blue photons of the line are actually
resonantly scattered by H~I, the red wing eventually escapes, leading to the
observed P~Cyg profile. They measured the velocity of expansion of the
superbubble deriving $\sim$$300$ km s$^{-1}$. They argued that the secondary
\lyalp\ emission could be produced at the leading front of the superbubble with
no neutral hydrogen ahead, explaining why this minor component is detected
blueshifted by $\sim$$300$ km s$^{-1}$ on top of the absorption trough caused by
the outflowing medium. Table~\ref{irastable} summarizes the basic properties of
the starburst galaxy IRAS~0833.

In this work we have analyzed HST, XMM-Newton, IUE and ground-based observations aiming to
characterize the properties of the starburst in IRAS~0833, and in order to check whether
the conclusions from the study of Haro~2 (also known as Mrk~33 and IRAS~10293+5439) by
\citet{Oti12} about the nature of the diffuse, extended \lyalp\ emission, could be
extended to another \lyalp\ emitting starburst galaxy with very different morphology and
global properties. We describe the observational data in Sect.~\ref{obsdatairas}, the
results are presented in Sect.~\ref{resultsiras}, and they are discussed in
Sect.~\ref{discussioniras}. Finally, conclusions are outlined in Sect.~\ref{conclusions}.

\begin{table*}
\caption[IRAS~0833 summary]{IRAS~0833 coordinates, redshift, distance, scale,
Galactic H~I, color excess toward the source, and oxygen abundance.}
\label{irastable}
\centering
\begin{tabular}{cccccccc}
\hline\hline\\
\multicolumn{1}{c}{R. A.} & \multicolumn{1}{c}{Decl.} & Redshift$^{\mathrm{a}}$ &
\multicolumn{1}{c}{Distance$^{\mathrm{a}}$} & \multicolumn{1}{c}{Scale$^{\mathrm{a}}$} &
\multicolumn{1}{c}{$N($H\,I$)^{\mathrm{b}}_{\rm{Gal}}$} & \ebv$_{\rm{Gal}}$ &
$12+\rm{log}(\rm{O}/\rm{H})^{\mathrm{c}}$ \\
(J2000.0) & (J2000.0) & & (Mpc) & (pc arcsec$^{-1}$) & (cm$^{-2}$) & & \\
\hline\\
$08$ $38$ $23.18$ & $+65$ $07$ $15.20$ & $0.0191$ & $80.2$ & $390$ & $4.5\times10^{20}$ &
$0.094$ &
$8.45$ \\
\hline
\end{tabular}
\begin{flushleft}
$^{\mathrm{a}}$ From NED (the NASA/IPAC Extragalactic Database).\\
$^{\mathrm{b}}$ Average value from \citet{Dickey90} and \citet{Kalberla05}.\\
$^{\mathrm{c}}$ Value from \citet{LopezSanchez06}.
\end{flushleft}
\end{table*}

\section{Observational data and synthesis models} 

\label{obsdatairas}

\subsection{HST observations: ultraviolet and optical}

We have used HST observations of IRAS~0833 obtained with ACS, STIS and GHRS, in the UV and
optical. In this section we describe the data and the processing done.

\subsubsection{HST images}

The UV, optical and \hal\ images used in this work were obtained and processed
by \citet{Ostlin09} from HST/ACS observations. The UV image we will use to study
the distribution of the massive, young stellar clusters corresponds to the
observation with the F140LP filter (PI: Kunth). On the other hand, the \hal\
image was created from the observation with the narrow filter FR656N, estimating
the \hal\ continuum by fitting pixel by pixel UV-optical photometric data points
(HST/ACS F140LP, F220W, F330W, F435W and F550M observations) with SED's obtained
from Starburst99 synthesis models \citep{Leitherer99}. The resulting continuum
was subtracted from the FR656N image. For a more complete description of the
methodology, see \citet{Ostlin09}. We show in Fig.~\ref{figirasuv} (left panel)
a multispectral image created by \'Angel R. L\'opez-S\'anchez (CSIRO/ATNF)
combining the HST/ACS images obtained with the F300W, F435W and FR656N filters.
In the right panel, the regions which we will analyze in the paper are labelled
on top of the F140LP image: 1) the spatially integrated emission, which
comprises the total radiation from the whole starburst in IRAS~0833 (STB), and
2) the central supermassive stellar cluster (SSC).

\subsubsection{HST spectroscopy}

We have reanalyzed in this work the low-resolution GHRS spectroscopic data of
IRAS~0833 described by \citet{GonzalezDelgado98} as well as the high resolution,
long-slit HST/STIS observations obtained by \citet{MasHesse03}. Previous high
resolution HST/GHRS observations by \citet{Kunth98} were not used, since the
HST/STIS data provided a broader wavelength coverage and higher spatial
resolution along the long slit. The GHRS observation was performed with a
squared aperture $1.7\arcsec \times 1.7\arcsec$ centered on the maximum of the
UV emission of IRAS~0833, i.e. its nucleus. Given its limited size, the aperture
covered only the region around the central supermassive stellar cluster (SSC).
The spectral coverage of the spectrum was $1100-1900$ \AA, which thus includes
the \lyalp\ line as well as the stellar absorption lines \siivd\ and \civd. The
log of the HST/GHRS observations is shown in Table~\ref{irasiuetable}.

The G140M and G430L gratings were used in the STIS observations of the source. A
$52\arcsec \times 0.5\arcsec$ slit was placed onto the nucleus of IRAS~0833 with
a position angle of $171\degr$. High spectral resolution data of the \lyalp\
line and the UV continuum in the spectral range $1200-1250$ \AA\ were obtained
with G140M. The G430L spectral image includes the emission of the nebular lines
\oxiil, \hb\ and \oxiiid. However, the integration time for G430L was very low,
and the observation was severely affected by cosmic rays. The extraction and
background removal processes in the STIS spectral images were carried out using
IRAF\footnote{IRAF is distributed by the National Optical Astronomy Observatory,
which is operated by the Association of Universities for Research in Astronomy
(AURA) under cooperative agreement with the National Science Foundation.},
following the procedure described by \citet{Oti12} for the STIS observation
along the minor axis of the starburst galaxy Haro~2. See \citet{MasHesse03} for
a more detailed description. The observation journal is included in
Table~\ref{irasstistable}. Figure~\ref{figirasaper} illustrates the position of
the GHRS aperture and the STIS slit onto the ACS F140LP UV image of IRAS~0833.

\begin{table*}
\caption[Log of HST/GHRS and IUE observations of IRAS~0833]{Log of HST/GHRS and IUE
observations of IRAS~0833 used in this work.}
\label{irasiuetable}
\centering
\resizebox{\textwidth}{!}{
\begin{tabular}{lcccccccc}
\hline\hline\\
Instrument & PI & Observation date & Grating/Camera &
\multicolumn{1}{c}{Integration time} &
\multicolumn{1}{c}{Aperture size} &\multicolumn{1}{c}{Position angle} &
\multicolumn{1}{c}{Wavelength interval} & \multicolumn{1}{c}{Spectral dispersion} \\
& & &  & (s)  & & (deg) & (\AA) & (\AA\ pixel$^{-1}$) \\
\hline\\
HST/GHRS & Robert & $1995$ Dec $31$ & G140L & $18088$ &
$1.7\arcsec\times1.7\arcsec$ & - & $1100-1900$ & $0.573$ \\ 
IUE & Thuan & $1989$ Feb $13$ & SWP & $4920$ & $20\arcsec\times10\arcsec$
& $51$ & $1150-1980$ & $1.68$ \\ 
IUE & Thuan & $1989$ Feb $13$ & LWP & $18000$ & $20\arcsec\times10\arcsec$
& $51$ & $1850-3350$ & $2.67$ \\ 
\hline
\end{tabular}
}
\end{table*}

\begin{table*}
\caption[Log of HST/STIS observations of IRAS~0833]{Log of HST/STIS observations of
IRAS~0833 used in this work.}
\label{irasstistable}
\centering
\resizebox{\textwidth}{!}{
\begin{tabular}{lccccccccc}
\hline\hline\\
Observation date & PI & Grating & Detector &
\multicolumn{1}{c}{Integration time} & \multicolumn{1}{c}{Position angle} &
\multicolumn{1}{c}{Slit size}  & \multicolumn{1}{c}{Wavelength interval} &
\multicolumn{1}{c}{Plate scale} & \multicolumn{1}{c}{Spectral dispersion} \\
& & & & (s) & (deg) &  & (\AA) & (arcsec pixel$^{-1}$) & (\AA\
pixel$^{-1}$) \\
\hline\\
$2001$ Jan $15$ & Kunth & G140M & MAMA & $7320$ & $171$ &
$52\arcsec\times0.5\arcsec$ & $1200-1250$ & $0.029$ & $0.053$ \\ 
$2001$ Jan $15$ & Kunth & G430L & CCD  & $360$ & $171$ &
$52\arcsec\times0.5\arcsec$ & $2900-5700$ & $0.050$ & $2.746$ \\ 
\hline
\end{tabular}
}
\end{table*}

\begin{table*}
\caption[Log of WHT/ISIS observations of IRAS~0833]{Log of WHT/ISIS observations of
IRAS~0833 used in this work.}
\label{irasisistable}
\centering
\resizebox{\textwidth}{!}{
\begin{tabular}{lccccccccc}
\hline\hline\\
Observation date & PI & Arm & Grating & \multicolumn{1}{c}{Integration
time} &
\multicolumn{1}{c}{Position angle} & \multicolumn{1}{c}{Slit size}  &
\multicolumn{1}{c}{Wavelength interval} & \multicolumn{1}{c}{Plate scale} &
\multicolumn{1}{c}{Spectral dispersion} \\
& & & & (s) & (deg) & & (\AA) & (arcsec pixel$^{-1}$) &
(\AA\ pixel$^{-1}$) \\
\hline\\
$2006$ Dec $25$ & L\'{o}pez & Blue arm & R600B & $3600$ &  $171$ &
$3.3\arcmin\times1.0\arcsec$ & $3460-5280$ & $0.20$ & $0.45$ \\ 
$2006$ Dec $25$ & L\'{o}pez & Red arm & R600R & $3600$ &  $171$ &
$3.3\arcmin\times1.0\arcsec$ & $5690-7745$ & $0.22$ & $0.49$ \\ 
\hline
\end{tabular}
}
\end{table*}

\subsection{IUE observations}

The IUE/SWP and LWP spectra of IRAS~0833 were downloaded from the IUE Newly
Extracted Spectra (INES) archive at CAB (Centre of Astrobiology, Madrid, {\tt
http://sdc.cab.inta-csic.es/ines/}). They were obtained with the IUE large
aperture ($20\arcsec \times 10\arcsec$), which enclosed most of the UV emission
of the source. The spectral coverage of the cameras are $1150-1980$ \AA\ (SWP)
and $1850-3350$ \AA\ (LWP). In Fig.~\ref{figirasaper} the position of its Large
Aperture relative to the ACS F140LP UV image of IRAS~0833 is shown. The log of
the IUE observation is shown in Table~\ref{irasiuetable}.

\subsection{XMM-Newton: X-rays}

\label{irasobsxrays}

\begin{figure*}
\centering
\includegraphics[width=8 cm,bb=0 0 1088 934
dpi,clip=true]{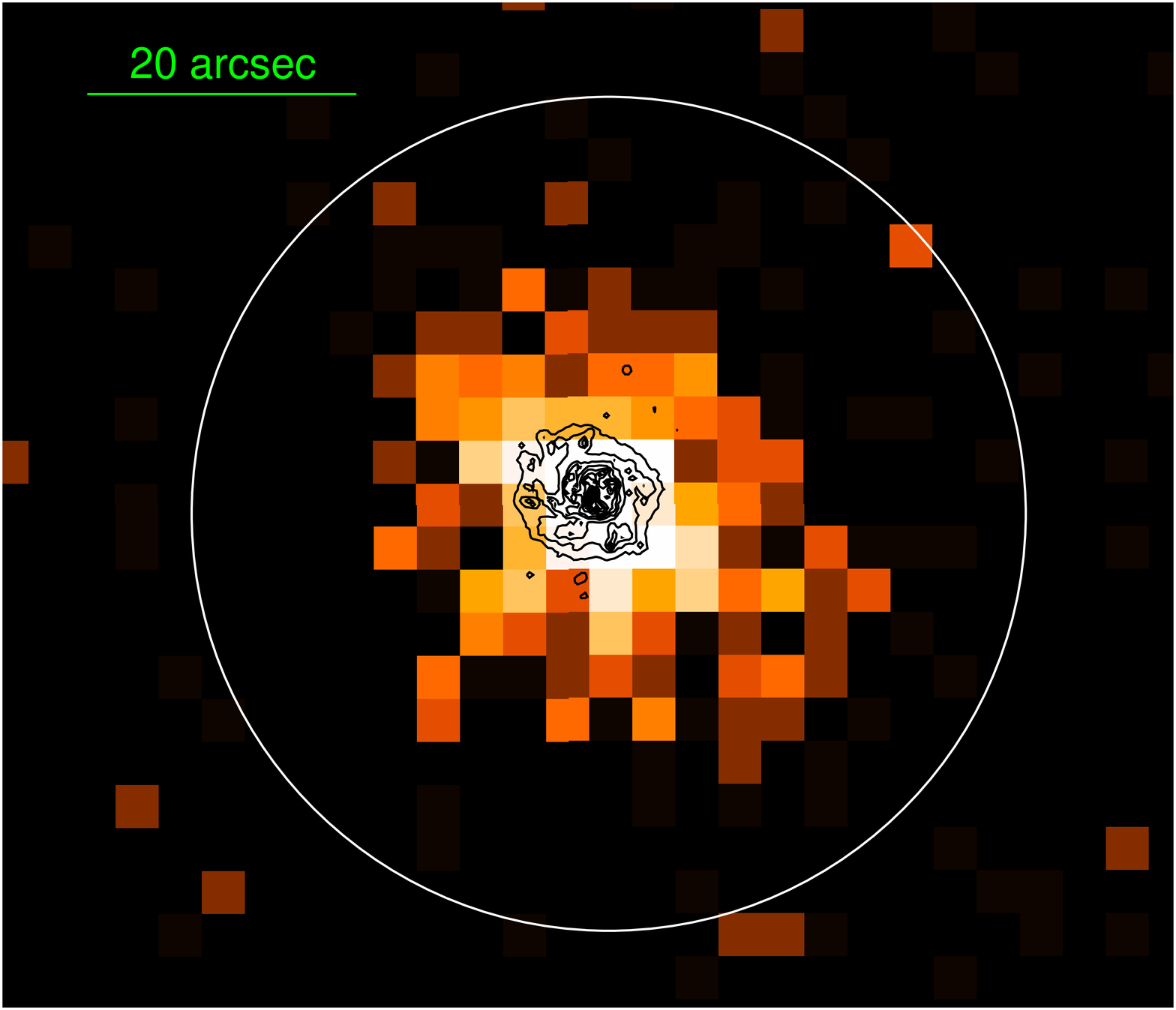}
\hspace{0.75 cm}
\includegraphics[width=8 cm,bb=0 0 1088 934
dpi,clip=true]{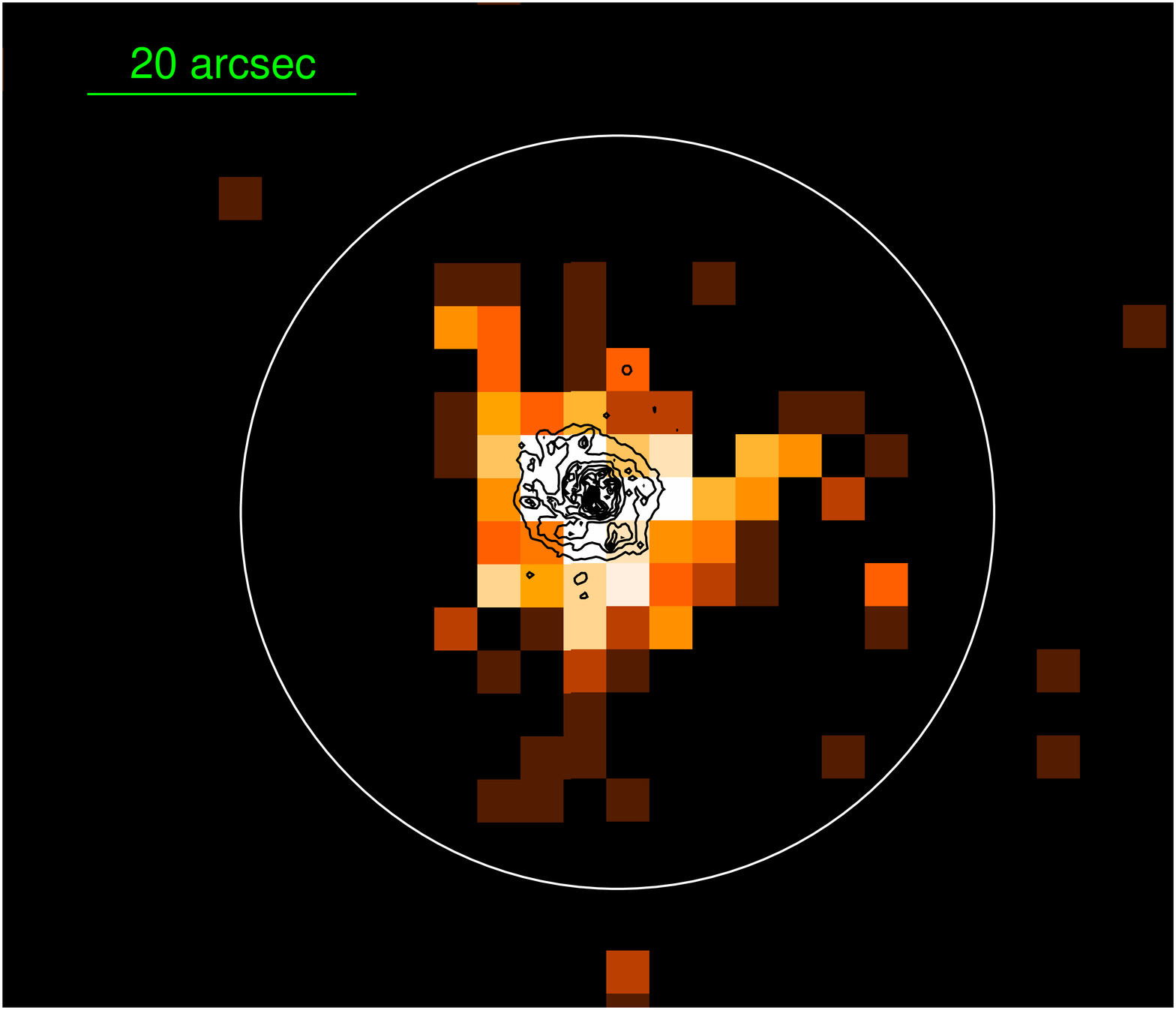}
\caption[X-ray image of IRAS~0833]{XMM-Newton/EPIC MOS1 (left) and MOS2 (right)
X-ray images of IRAS~0833. Intensity scale is logarithmic. Source region from
which X-ray spectrum was extracted for each camera is confined within the white
circle. Contours from HST/ACS F144LP image are shown in black. PSF of this
off-axis observation is large and asymmetric, and thus the emission must be considered as
non-resolved.
North is up and east is left.} 
\label{figirasx}
\end{figure*}

XMM-Newton observation $0111400101$ of $\pi^{1}$ Ursae Majoris (PI: Albert
Brinkman) included IRAS~0833 in its field of view $\sim 8\arcmin$ away from the
optical axis. Data from XMM-Newton analyzed in this work were obtained by the
European Photon Imaging Camera (EPIC), which consists of 2 MOS (Metal Oxide
Semi-conductor) and 1 pn CCD arrays. Each detector is located in a different
X-ray telescope on-board XMM-Newton, and whereas only around half of the
radiation entering the corresponding telescopes reaches the MOS cameras, the pn
camera receives an unobstructed beam. The EPIC observation consisted in imaging
of the source with Full Frame (MOS1) and Small Window (MOS2), as well as one
timing observation (pn). We focused on the MOS1 and MOS2 imaging observations
since we are not interested in the timing analysis of the source. Total exposure
time of the observation was $50.1$ ks (MOS1) and $45.2$ ks (MOS2). Reduction of
the data was performed with the Scientific Analysis Software 13.0.0 (SAS),
following the standard threads available in the website of the XMM-Newton-SAS.
Only spectral range $0.3-10.0$ keV was considered, as well as events with
patterns\footnote{{\tt PATTERN} describes how the charge cloud released by the
incoming X-ray photon distributes over the pixels. See the SAS manual for a more
detailed description.} {\tt PATTERN}~$\leq 4$ in order to maximize the
signal-to-noise ratio (SNR) against non-X-rays events. Source regions in the
EPIC images were calculated with SAS task {\tt eregionanalyse} in order to
increase the SNR value over the background, resulting in circular regions with
{\em radius} of $31\arcsec$ (MOS1) and $28\arcsec$ (MOS2). Redistribution
matrices and ancillary files were calculated for the MOS1 and MOS2 images
assuming the source regions for each image from which spectrum was extracted.
Figure~\ref{figirasx} shows the MOS1 and MOS2 images, together with the circular
source regions assumed and the UV countours from the HST/ACS F140LP image.
Source in MOS2 happened to be located close to a CCD border, and thus a fraction
of the total X-ray emission may have been missed in this image. For consistency,
when modeling the total spectrum we checked that similar results were obtained
for both MOS spectra. Background was extracted from two circular regions with
{\em radius}~$\sim 1\arcmin$ with different locations in each of the two images
but which would be close to IRAS~0833 and would not include any bright source. A
net count rate of $\sim$$1.5\times10^{-2}$ cts s$^{-1}$ was measured in the
source regions after background subtraction. We used {\em XSPEC} v.12.8
\citep{Arnaud96} to analyze the spectral data. Source counts were grouped to
have at least 25 counts per bin in order to be able to apply $\chi^{2}$
statistics in the fitting analysis. The final binned X-ray spectrum of IRAS~0833
is shown in Fig.~\ref{figirasxspec}. Since this is an off-axis observation, the
point spread function (PSF) of the instrument is large and distorted. We
compared the radial profiles of the source with the EPIC PSF using SAS task {\tt
eradial} at low, medium and high energy, and concluded that IRAS~0833 is not
resolved and can be treated as a point-like source. Therefore, no information on
the structure of the X-ray-emitting regions is available in Fig.~\ref{figirasx},
but just the spectral properties of their integrated emission.

\begin{figure*}
\begin{center}
\includegraphics[height=8 cm,bb=0 0 534 515
dpi,clip=true]{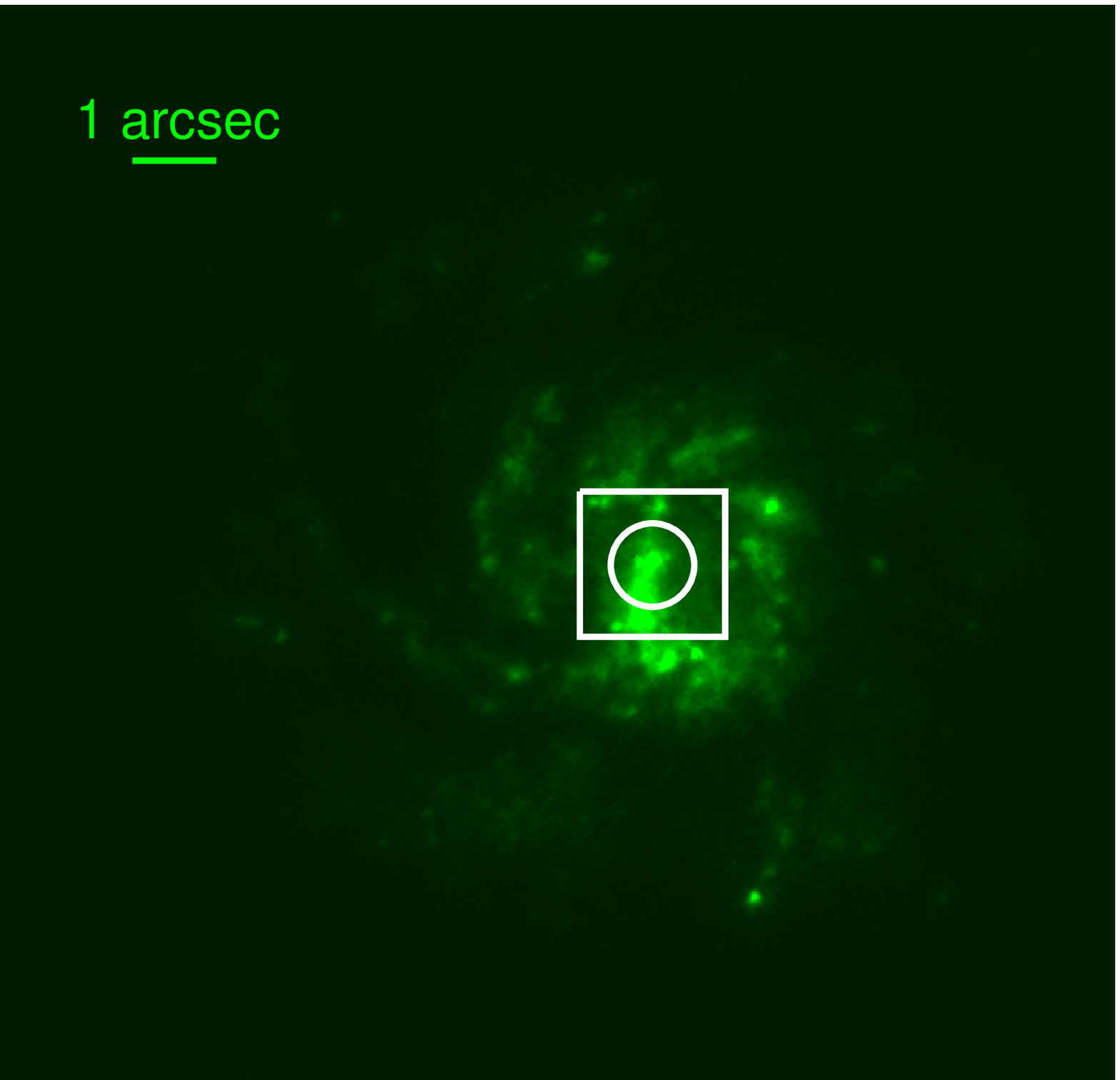}
\hspace{0.75 cm}
\includegraphics[height=8 cm,bb=0 0 534 515
dpi,clip=true]{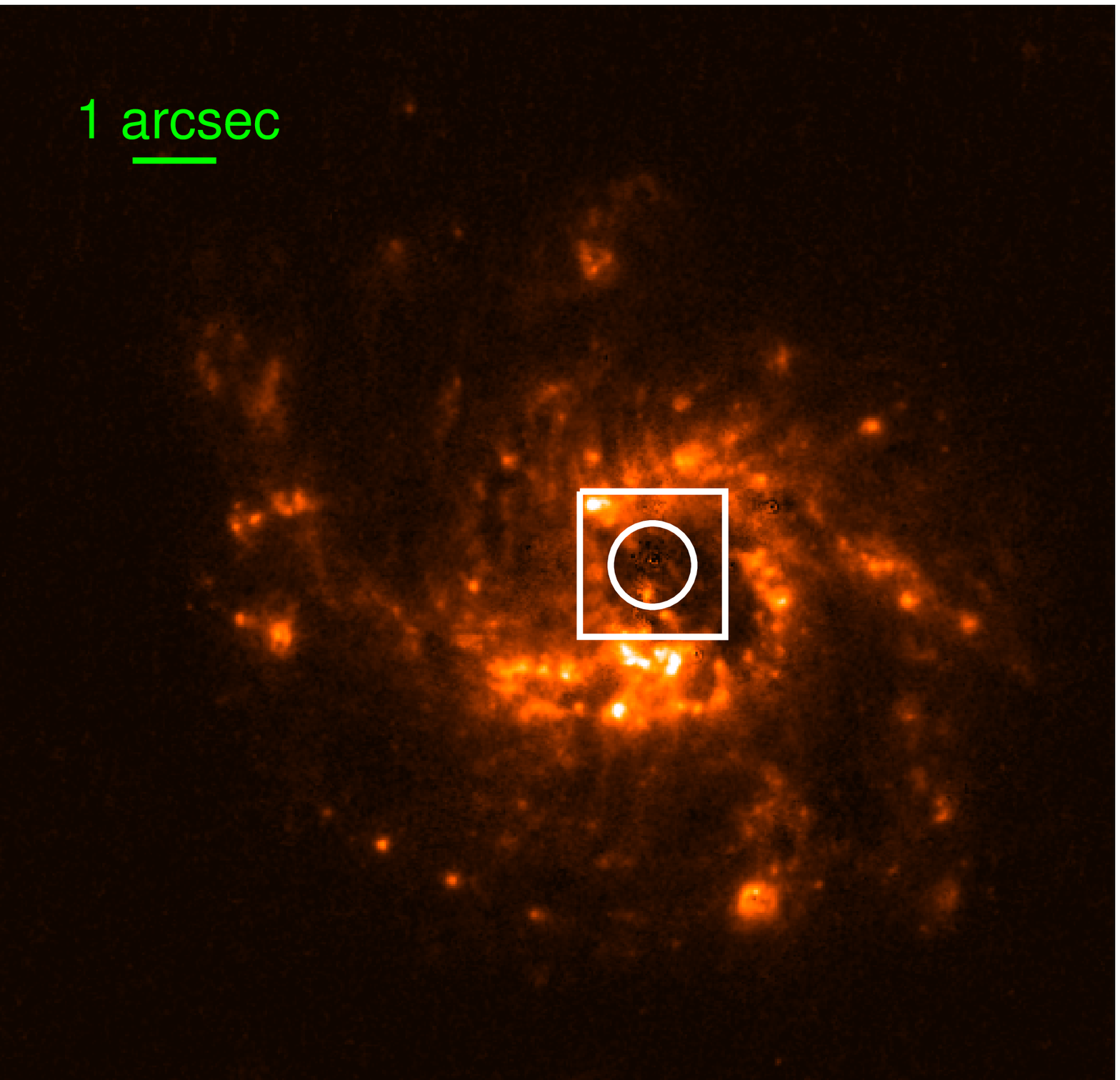}
\vspace{0.5 cm}
\\
\includegraphics[height=8 cm,bb=0 0 534 515
dpi,clip=true]{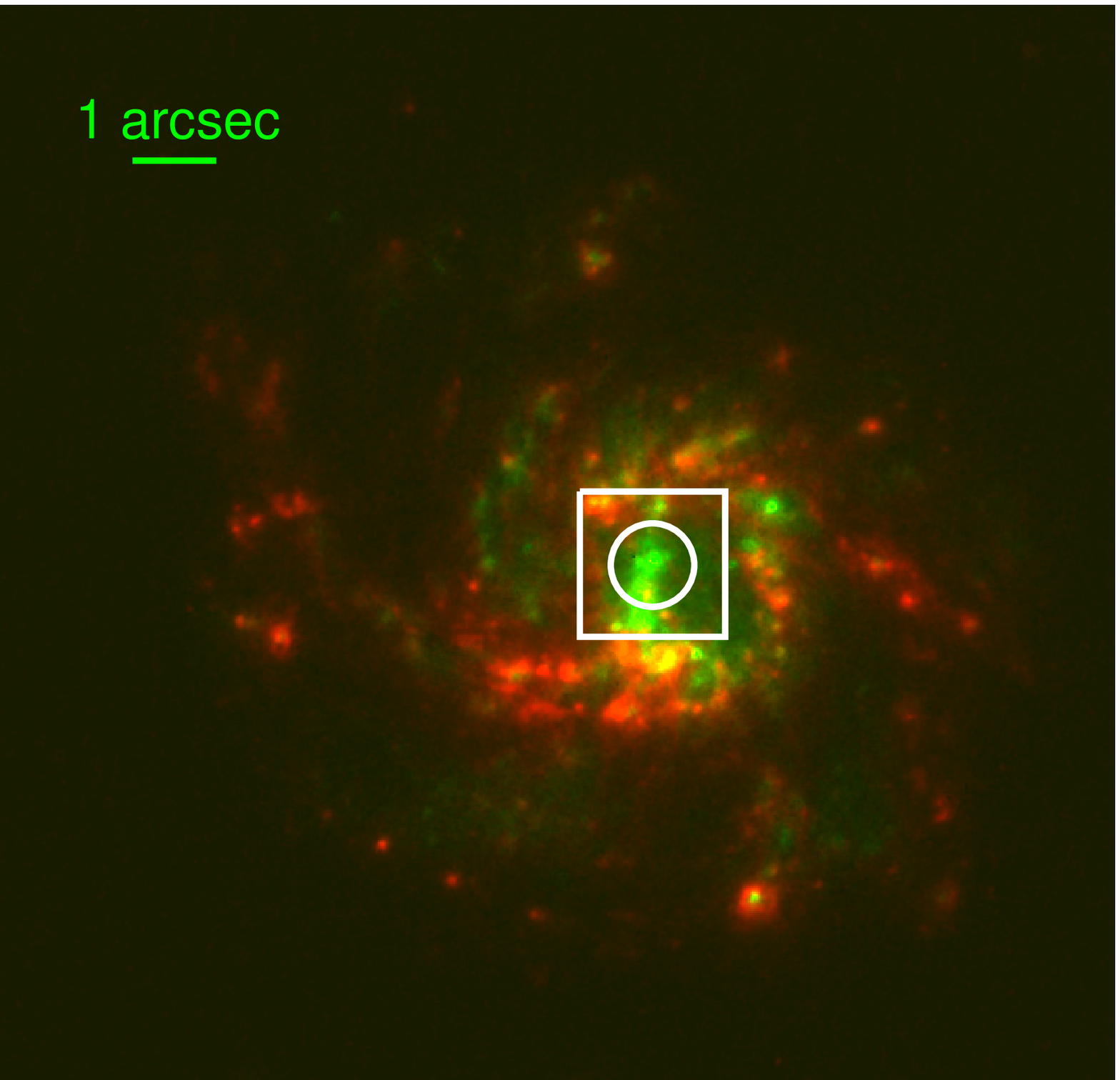}
\hspace{0.75cm}
\includegraphics[height=8 cm,bb=15 15 600 554
dpi,clip=true]{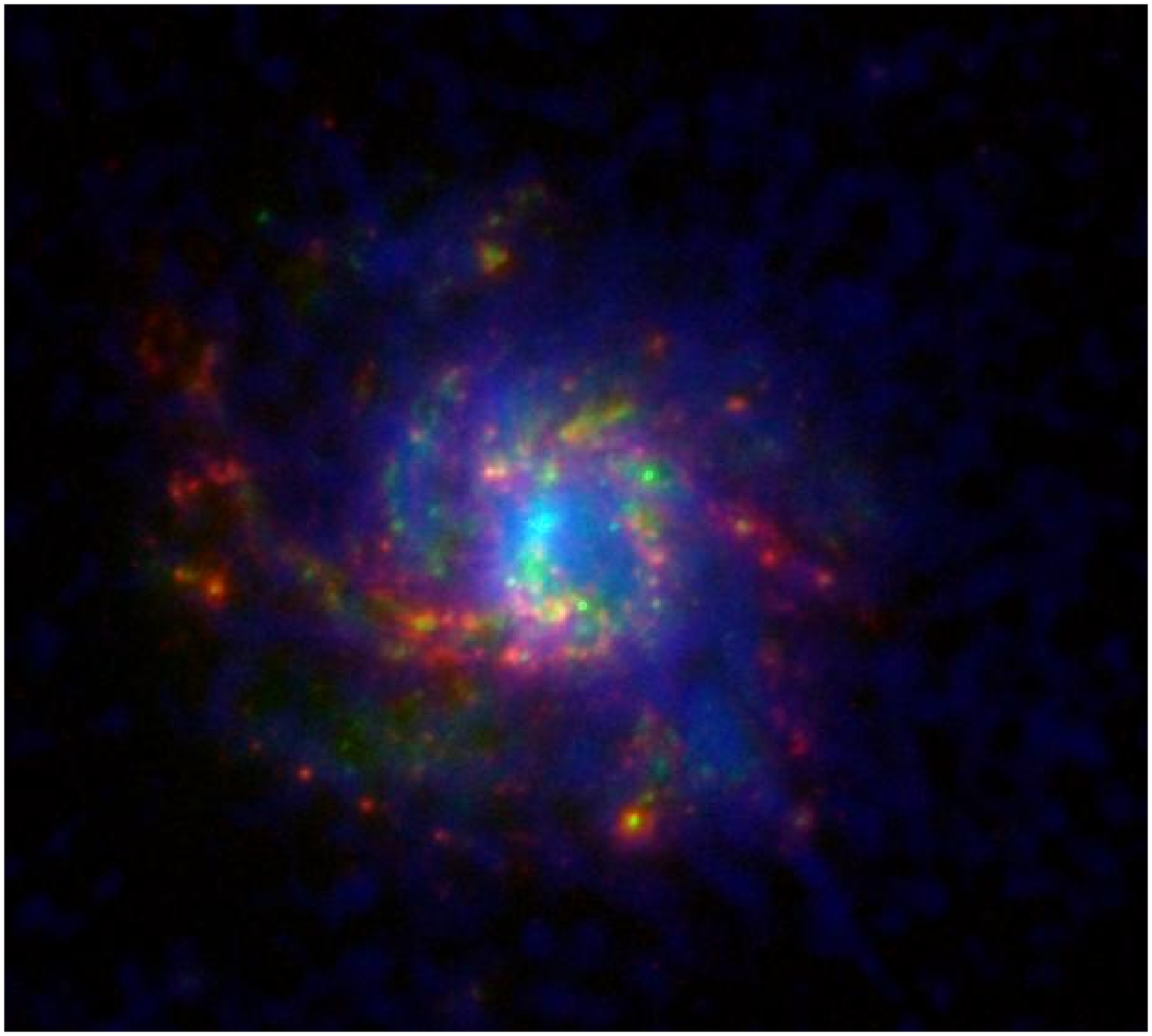}
\end{center}
\caption[UV-\hal-\lyalp\ color-coded images of IRAS~0833]{HST/ACS images of
IRAS~0833 in the ultraviolet (filter F140LP, green, {\em top left}),
continuum-subtracted \hal\ (red, {\em top right}), the composite of both ({\em
bottom left}), and the composite including \lyalp\ emission in blue ({\em bottom
right}, adapted from \citet{Ostlin09}). The aperture of HST/GHRS is marked with
a square. The circular region $R_1$ ({\em radius}~$\sim 0.5\arcsec$ ($\sim$$200$
pc)), as discussed in the text, is also marked. Intensity scale is linear,
except for the last image, which is logarithmic. North is up and east is left.}
\label{figirasuvha}
\end{figure*}

\subsection{Ground-based observations}

Two long-slit observations with the blue and red arms of William Herschel Telescope/ISIS
were downloaded from the archive (PI: \'Angel R. L\'opez-S\'anchez). The
location and position angle of the $1.0\arcsec$-wide ISIS slit coincided with that of the
HST/STIS slit. The spatial profiles of the \hal\ and \hb\ emission lines were extracted
from the ISIS spectral images. The log of these observations is included in
Table~\ref{irasisistable}.

\subsection{Synthesis models}

Evolutionary population synthesis models by \citet{Leitherer99} (Starburst 99,
herineafter SB99) and \citet{Cervino02b} (hereinafter CMHK02) have been used to
characterize the star formation episode taking place in IRAS~0833. Both sets of
models compute the evolution of a cluster of massive stars for a given initial
mass function, metallicity and star formation regime. Once the structure (number
of stars of each spectral type and luminosity class at a given evolutionary
time) of the stellar population is derived, the models are used to calculate a
set of observable parameters (continuum emission, number of ionizing photons
produced, supernova rate, equivalent width and intensity of nebular lines,
etc.). Two different star formation regimes are considered in these models: 1)
stars are produced at a constant rate (extended burst, EB), and 2) prompt
formation occurs during a short period of time, after which no more stars are
produced (instantaneous burst, IB). For consistency, we assume in both cases the
same Salpeter initial mass function (IMF, $\phi(m) \sim m^{-2.35}$) with mass
limits of $2-120$ \msun. While the lower mass stars do not produce any
observable effect in the UV range, they provide the bulk of the stellar mass and
can even dominate the optical--IR continuum after some tens of Myr of evolution.
The output magnitudes of the models are scaled by the intensity of star
formation, which is characterized by the star formation rate (SFR, the velocity
at which stars are being produced) in EB models, and star formation strength
(SFS, total mass converted into stars) in IB models. See the cited references
for a more detailed description of the models.

\section{Analysis and results} 

\label{resultsiras}

\subsection{HST morphology and photometry}

The F140LP UV image in Fig.~\ref{figirasuv} shows that IRAS~0833 is composed of
a large, bright central stellar cluster along an internal bar almost aligned
from N to S, surrounded by a rather diffuse UV emission from several spiral arms
which enclose smaller, unresolved stellar knots. Whereas the elongated central
cluster is comprised within $1\arcsec \times2\arcsec$ ($\sim 400$ pc $\times$
$800$ pc), the spiral arms extend over a circular region of {\em radius}~$\sim
5.5\arcsec$ ($\sim$$2$ kpc). As shown in the right panel of
Fig.~\ref{figirasuv}, hereinafter we will refer to the global starburst in
IRAS~0833 as STB, and we will characterize it by analyzing the integrated
emission of the whole source. On the other hand, we have labelled its central
UV-bright stellar cluster as SSC, whose emission is roughly included within the
GHRS aperture.

The observed total UV flux of the STB region (\fuviras) was measured at $1500$
\AA\ rest frame in two ways: 1) through the integrated emission of HST/ACS
F140LP image, and 2) from the flux at $1500 (1+z)$ \AA\ in the IUE spectrum.
Although very large, the IUE aperture might have not included the whole emission
from the source. To check whether this was the case, we calculated both the
total integrated flux from the HST/ACS F140LP image ($F_{\rm{UV\,
integrated}}$), as well as the flux enclosed in the same image by an aperture
similar to the IUE one positioned with the same angle ($F_{\rm{UV\, IUE}}$), as
shown in Fig.~\ref{figirasaper}. We derived a ratio $F_{\rm{UV\,
integrated}}/F_{\rm{UV\, IUE}}=1.04$, indicating that the IUE aperture encloses
$96$\% of the UV emission of IRAS~0833. After correcting for this aperture loss,
the value of the IUE UV flux remains $15$\% lower than the integrated HST/ACS
F140LP flux. The disagreement is most likely a combination of the $10$\%
photometric accuracy of IUE (see Fig.~\ref{figirassedall}) and the very broad,
skewed shape of the F140LP filter. We therefore decided to take the average of
both flux measurements as the total UV flux at $1500$ \AA\ of STB,
\fuviras~$=4.5\times 10^{-14}$ \ergsacm. This uncertainty in the measurement of
\fuviras\ translates into a systematic uncertainty of $\pm 10\%$ in the
determination of any absolute magnitude of STB based on it, like the total mass
of the starburst or the predicted luminosities at different ranges.

Given the size of its aperture, the spectrum obtained by HST/GHRS includes
mostly the emission from the central cluster, which we have labelled SSC. The UV
flux value found for this region \fuviras~$=8.5\times 10^{-15}$ \ergsacm\ is
$20$\% of STB, i.e. the total integrated observed UV flux of IRAS~0833.

We show in Fig.~\ref{figirasuvha} a color-coded image of IRAS~0833 combining F140LP
UV (green) and \hal\ (red), along with the single images of each filter, as
generated by \citet{Ostlin09}, and merged together using {\em ds9}
\citep{Joye03}. While in the \hal\ image  a filamentary structure extending
largely around the stellar clusters in the spiral arms appears very prominent,
the region dominated by the UV-bright SSC shows a very low
\hal\ emission, if any.

\subsection{UV stellar continuum}

\begin{figure*}
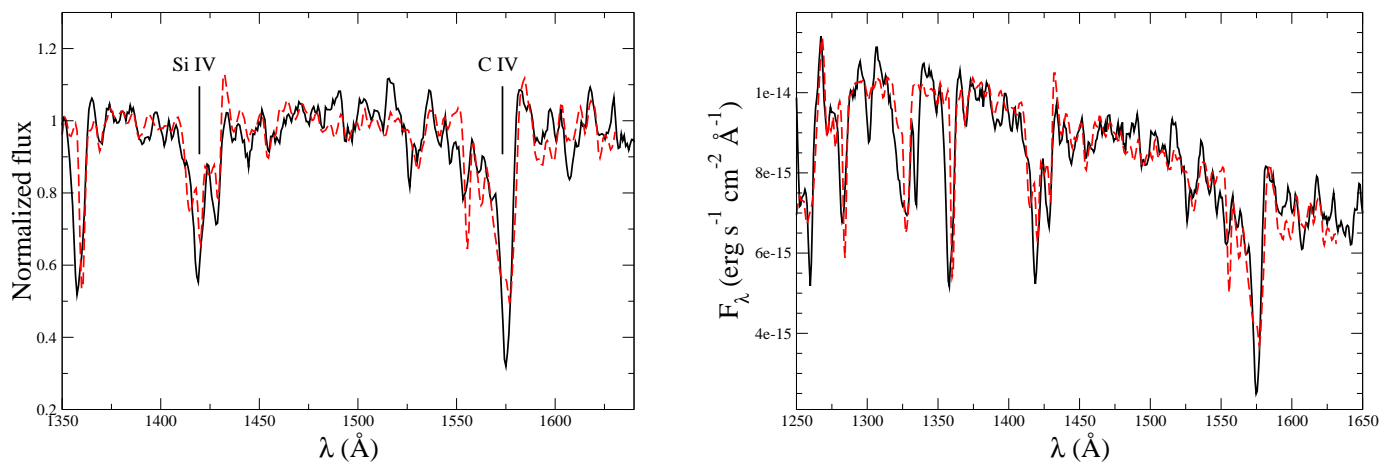

\centering
\includegraphics[height=6 cm,bb=37 34 706 522
dpi,clip=true]{iras0833.norm_spec.lines_s03.IB.07.eps}
\hspace{0.75 cm}
\includegraphics[height=6 cm,bb=2 34 722 522
dpi,clip=true]{fit_spectrum_GHRS.iras_0833.SB99_uvline.08.print.eps}
\caption[SED and normalized UV spectrum of IRAS~0833]{HST/GHRS spectrum of SSC
in IRAS~0833 (solid line) fitted with an SB99 IB model for an age of $\sim$$5.5$
Myr (dashed line). Left: Normalized spectrum with absorption lines \siivd\ and
\civd\ labelled. Right: regular spectrum reddened to account for Galactic
(Cardelli extinction law, \ebv=$0.094$) and intrinsic (LMC law, \ebv=$0.01$)
extinctions.}
\label{figirasnorm}
\end{figure*}

In order to estimate the evolutionary state of the star formation episode in
IRAS~0833, the profile of the spectral lines \siivd\ and \civd\ in the
normalized HST/GHRS (SSC) and IUE (STB) spectra were fitted using
high-resolution UV spectra generated with the SB99 models. As shown in
Fig.~\ref{figirasaper} and as already discussed above, whereas the HST/GHRS
spectrum only covers the central stellar cluster of IRAS~0833, the
low-resolution IUE spectrum contains most of the integrated UV emission of
IRAS~0833 (STB). Both the Milky Way and the LMC/SMC libraries were used in the
SB99 models, assuming a Salpeter IMF ($\phi(m) \sim m^{-2.35}$) with mass limits
of $2-120$ \msun\ and $Z=0.008$ for the evolutionary tracks and the
high-resolution spectra. The metallicity value considered was the closest
available to the observed one \citep{LopezSanchez06}. We started by fitting the
GHRS spectrum of the SSC, given that its higher spectral resolution allows a
more detailed analysis than the IUE one. We checked that no acceptable fitting
was possible for the C~IV lines with the Milky Way library. The model that
reproduces more accurately the spectral profile of the lines is an instantaneous
burst (IB) model of age $\sim$$5.5$ Myr obtained with the LMC/SMC library, in
agreement with the age range $4-6$ Myr found by \citet{LopezSanchez06}.
Furthermore, these authors detected the presence of WR stars within an aperture
similar to the HST/GHRS one, with a ratio $WR/(WR+O)\sim 0.03$. According to
evolutionary synthesis models \citep{Schaerer98,Cervino94} the presence of WR
stars at the metallicty of IRAS~0833 is only possible if the prompt star forming
event is younger than around 6~Myr. The measured $WR/(WR+O)$ ratio is consistent
with the predictions for a starburst at $5-5.5$ Myr \citep{Schaerer98}. Even
more, the $WR/(WR+O)$ value is incompatible with an extended starburst (EB)
which had been producing stars for more than 10~Myr, since then it would show
much lower values ($WR/(WR+O)< 0.009$ \citep{Cervino94}). Both the observational
HST/GHRS and the SB99 normalized spectra around the \siivd\ and \civd\ lines are
shown in Fig.~\ref{figirasnorm} (left panel).

Finally, we checked that the normalized IUE spectrum of STB is consistent  with
the age obtained from the GHRS data of the SSC region, as far as the lower IUE
spectral resolution allows. Though we can not exclude some age spread of the
star-forming knots along the spiral arms, we conclude that the global star
formation episode in IRAS~0833 is compatible with a short-lived burst which
started $\sim$$5.5$ Myr ago over the whole galaxy.

\subsection{Extinction}

\label{irasext}

We derived the average reddening of the integrated UV emission of IRAS~0833
(STB) by fitting the IUE spectrum with the SB99 model computed above. The fit
was performed by applying a certain intrinsic \ebv\ value to the predicted
continuum. The spectrum was then redshifted ($z=0.0191$) and finally reddened
again assuming \ebv~$=0.094$ and Cardelli extinction law \citep{Cardelli89} to
account for the Galactic extinction. The procedure was repeated for different
intrinsic \ebv\ values until the observed UV slope and shape were satisfactorily
reproduced. As can be seen in Fig.~\ref{figirassed} the IUE spectrum of
IRAS~0833 shows a prominent $2175$ \AA\ feature, typically associated in the
past to graphite in the dust grains, but whose origin might be rather in
hydrogenated amorphous carbon materials \citep{Gadallah11}. The SMC law
\citep{Prevot84} was not considered to model the intrinsic extinction since it
does not reproduce the $2175$ \AA\ bump, unlike the LMC \citep{Fitzpatrick85}
and Galactic laws \citep{Cardelli89}. We assumed $R_{V}=3.1$ for these laws.
Once the reddening correction was applied, we computed the mass of the
integrated source by dividing the observed UV flux at $\sim$$1529$ \AA\
($\sim$$1500$ \AA\ rest frame) by the reddened, redshifted value predicted by
the SB99 model. The best fit yielded \ebv~$=0.15$ assuming an LMC-like law and
$M \sim 1.4\times10^{8}$ \msun\ for the initial mass of stars (in the 2-120
\msun\ range) having formed all over IRAS~0833 around 5--6 Myr ago. This value
of the extinction is in agreement with \citet{GonzalezDelgado98} who obtained
\ebv~$=0.17$ with an LMC law when fitting the integrated continuum emission
observed with HUT (Hopkins Ultraviolet Telescope). When assuming the Cardelli
Galactic law for the internal extinction, either the $2175$ \AA\ feature or the
UV slope at $\lambda<1800$ \AA\ was reproduced, but not both simultaneously. As
shown in Fig.~\ref{figirassed} (left panel), the best fit is obtained using the
LMC law for the internal extinction, although the $2175$ \AA\ is not completely
reproduced. If Cardelli law is assumed, \ebv=$0.20$ would be required, but the
fits neither improve, nor reproduce correctly the slope of the UV continuum.
\citet{Leitherer13} measured \ebv$=0.25$ when fitting the spectrum of IRAS~0833
obtained with HST/COS (circular aperture with {\em radius}$=1.25\arcsec$), which
only includes the spectral range $1100-1450$ \AA. This fit is significantly
worse than the one achieved with the LMC law and \ebv=$0.15$ when the whole UV
range is considered. We want to stress that, contrary to the optical range,
different  E(B-V) values are required to get the same extinction factor in the
UV for different laws.

The reddening affecting the UV continuum of the central stellar cluster in
IRAS~0833 (SSC) was also computed by fitting the HST/GHRS spectrum with the
predicted SB99 SED, obtaining a very low extinction with \ebv~$=0.01$. We
assumed an LMC-like extinction law, as derived from the IUE spectrum. The
observed UV continuum, together with the best fitting, reddened SB99 model, have
been plotted in Fig.~\ref{figirasnorm} (right panel). Assuming this extinction,
the UV continuum emitted by the SSC is compatible with a burst at $5.5$ Myr with
an initial mass of $M$~$\sim 7.0\times10^{6}$ \msun, within the assumed mass
limits of $2-120$ \msun.

The values for the best fits are given in Tables~\ref{irasinttable} and
\ref{irasssctable}, including the reddening and mass values for both the STB and
SSC regions. Also, the ratios of the UV and \hal\ emissions and the masses of
SSC over STB, i.e. the central region over the integrated IRAS~0833, are
included. The compact nature and the mass value of the central stellar cluster
in IRAS~0833 indicates that it must be considered a Super Stellar Cluster
\citep{Adamo10}, hence its name SSC. As shown in Table~\ref{irasssctable},
whereas the SSC contributes up to $20$\% to the total observed UV flux, its
intrinsic UV emission accounts for only $5$\% of the integrated unreddened
value. This means that around $5$\% of the mass within STB is associated  to the
SSC. Finally, in Fig.~\ref{figirassedall} we show the HST/GHRS spectrum,
together with the IUE one. Also, the predicted SB99 model ($age=5.5$ Myr,
$Z=0.008$) is included, which has been reddened only by the Galactic extinction
(\ebv~$=0.094$). As can be seen, the stellar continuum from the SSC is hardly
affected by dust extinction, showing a slope similar to the unreddened SB99
model, unlike the STB continuum, which actually shows a positive UV slope below
1500~\AA.

\subsubsection{Differential extinction}

Using NOT/ALFOSC, \citet{LopezSanchez06} obtained an \hal\ image of IRAS~0833,
as well as a spectral image using ALFOSC in spectrographic mode with a
$6.4\arcmin \times1\arcsec$ slit placed on the center of the source and with a
position angle PA~$=138\degr$. Since the spectral range extended from 3200 to
6800 \AA, \citet{LopezSanchez06} could calculate the Balmer decrement and hence
the nebular dust extinction. They obtained an integrated value of \chb~$=0.22$
for the nebular extinction (see \citet{Mazzarella93} for the definition of
\chb), which corresponds to \hal$/$\hb~$=3.4$ including also the Galactic
contribution. This translates into an internal nebular extinction of
\ebv$=0.06$. On the other hand, these authors found that the nebular extinction
is largest within the central region $\sim$$1\arcsec\times1\arcsec$ ($\sim$$400$
pc $\times$ $400$ pc), with \chb~$=0.30$ (corresponding to \ebv$\sim0.11$ after
correction from Galactic extinction).

The effect is reversed when we look at the extinction affecting the stellar
continuum. As shown in Fig.~\ref{figirassedall}, while the integrated stellar UV
continuum from STB is severely affected by dust, with a global \ebv$\sim0.15$
(corrected from Galactic extinction), the stellar continuum from SSC shows a
very low, almost negligible extinction (\ebv$\sim0.01$).

Furthermore, Fig.~\ref{figirasuvha} shows that the \hal\ emission is
concentrated in circumnuclear blobs around the central cluster and in the spiral
arms, but is very weak just on top of it. The origin of these apparent
discrepancies is related to the cavity apparently devoid of gas and dust which
surrounds the central SSC. The mechanical energy released by the SSC in the
form of stellar winds and supernova explosions has efficiently swept out
the gas and dust, leaving a rather clean region just around the central SSC. 
This explains the low reddening affecting the UV continuum and the lack of
\hal\ emission in the close vicinity of the SSC. These stellar winds have
apparently concentrated the dust and gas at larger distances (but still close to
the SSC), so that the \hal\ emission integrated over this central region appears
severely obscured. Outside this central region, the distribution of dust is
smoother. In Fig.~\ref{figirasratios} the spatial profile of \hal/\hb\ as
derived from the WHT/ISIS observations is shown to be quite uniform over large
areas, with just a weak gradient from north to south. This scenario exemplifies
the spatial decoupling between the extinction affecting the stellar
continuum and that affecting the nebular lines, which is very frequent in
starburst galaxies and can yield misleading conclusions when analyzed
globally (see further discussion below in Sect.\ref{irasdisccentral}).

\subsection{Far infrared emission}

Once the starburst in IRAS~0833 had been characterized and its main properties
constrained, we checked whether the FIR and \hal\ luminosities from STB could
also be reproduced. For this task we used the evolutionary synthesis models by
CMHK02 models, which allow to compute the expected FIR flux for a given burst
and a specific extinction. \citet{LopezSanchez06} reported an integrated \hal\
luminosity \lha~$=1.2\times10^{42}$ \ergs\ already corrected for N~[II]
contamination and internal and Galactic extinctions. Assuming the starburst
model described  above, the expected intrinsic \hal\ emission predicted by
CMHK02 models for STB in IRAS~0833 is \lha~$=1.8\times10^{42}$ \ergs, which is
around $50$\% higher than the observed value.

\begin{figure*}
\centering
\includegraphics[height=5.9 cm,bb=1 34 706 522
dpi,clip=true]{fit_spectrum_IUE.iras_0833.SB99_spectrum.02.print.eps}
\hspace{0.75 cm}
\includegraphics[height=5.9 cm,bb=1 34 722 522
dpi,clip=true]{SED.iras_0833.SB99_spectrum.02.print.eps}
\caption[Fitting of the integrated SED of IRAS~0833]{{\bf Left:} IUE spectrum
of STB in IRAS~0833 (black, solid line) fitted with an SB99 spectrum for
an age of $5.5$ Myr (dashed lines). The model has been reddened to account for
the Galactic extinction, as well as for the intrinsic one assuming both an LMC
law and \ebv=$0.15$ (red), and the Cardelli extinction law with \ebv=$0.20$
(green). {\bf Right:} IUE spectrum is shown as in the left panel (black,
solid line). A composite model (blue, dashed line) was obtained summing the
starburst model (red) and an SB99 spectrum of a 1-Gyr-old population reddened
similarly as the young burst model (magenta, dashed line). Near-UV and optical
photometric fluxes from \citet{Ostlin09} are marked with green crosses.}
\label{figirassed}
\end{figure*}

\begin{figure}
\centering
\includegraphics[width=9 cm,bb=5 34 706 522
dpi,clip=true]{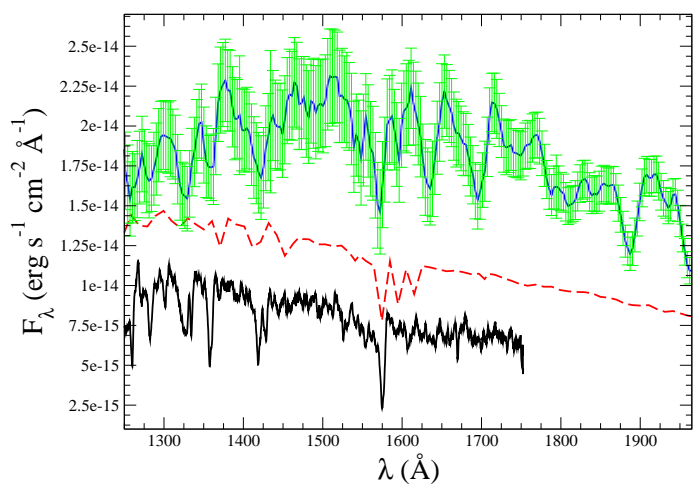}
\caption[Integrated and nuclear SEDs of IRAS~0833]{Observed HST/GHRS (black,
solid line, SSC) and IUE (blue, solid, STB; errors in green) spectra, together
with the SB99 model for an age of $5.5$ Myr (red, dashed). Vertical scale
corresponds to the GHRS spectrum. The other spectra have been arbitrarily scaled
for an easier comparison. SB99 model has been corrected only for Galactic
extinction.}
\label{figirassedall}
\end{figure}

\begin{table*}
\caption[Physical properties of the starburst in IRAS~0833]{Predicted values of
the age, mass, stellar and nebular extinctions, fraction of ionizing photons
absorbed by dust, emission fluxes and luminosities and \ewhb\ of the starburst
(STB) in IRAS~0833, assuming the UV fluxes from HST/ACS and IUE and the average
value of both. The Galactic extinction effect has been removed from the \ebv\
values, which thus strictly correspond to the intrinsic extinction within the
source. Contribution to the optical continuum by old stars was considered when
estimating \ewhb\ (see Sect.~\ref{irasunder} for details). Observed values are
also shown.}
\label{irasinttable}
\centering
\resizebox{\textwidth}{!}{
\begin{tabular}{lcccccccccc}
\hline\hline\\
& $Age$ & $M$ & \ebv$_{*}$ & \ebvneb\ & $1-f$ & \fuviras\ &
\lcuva$^{\mathrm{a}}$ & \lha$^{\mathrm{a}}$ & \lfir\ & \ewhb\  \\
& (Myr) & (\msun) &  &  & & (\ergsacm) & (\ergsa) & (\ergs) & (\ergs) & (\AA) \\
\hline\\
Model(ACS) & $5.5$ & $1.5\times 10^{8}$ & $0.15$ & $0.06$ & $0.5$ & $4.8\times
10^{-14}$ & $3.2\times 10^{41}$ & $1.3\times 10^{42}$ & $4.5\times 10^{44}$ &
$14$ \\ 
Model(IUE) & $5.5$ & $1.3\times 10^{8}$ & $0.15$ & $0.06$ & $0.5$ & $4.1\times
10^{-14}$ & $2.7\times 10^{41}$ & $1.2\times 10^{42}$ & $3.8\times 10^{44}$ &
$14$ \\ 
Model(average) & $5.5$ & $1.4\times 10^{8}$ & $0.15$ & $0.06$ & $0.5$ &
$4.5\times 10^{-14}$ & $3.0\times 10^{41}$ & $1.2\times 10^{42}$ & $4.2\times
10^{44}$ & $14$ \\ 
\hline\\
Observed & - & - & $0.17^{\mathrm{b}}$ & $0.06^{\mathrm{c}}$ & - & $4.5\times
10^{-14}$ & $3.0\times 10^{41}$ & $1.2\times 10^{42}$ $^{\mathrm{c}}$ &
$3.6\times 10^{44}$ & $19^{\mathrm{c}}$ \\  
\hline
\end{tabular}
}
\begin{flushleft}
$^{\mathrm{a}}$ Luminosity value corrected for Galactic and intrinsic extinctions.\\
$^{\mathrm{b}}$ Value from \citet{GonzalezDelgado98}.\\
$^{\mathrm{c}}$ Value from \citet{LopezSanchez06}.
\end{flushleft}
\end{table*}

\begin{table*}
\caption[Integrated (STB) and nuclear (SSC) physical properties of
IRAS~0833]{Observed values of the UV and \hal\ emissions and the stellar and
nebular extinctions of the integrated source IRAS~0833 (STB) and its central
stellar cluster (SSC), together with the predicted stellar mass for each of
them. The Galactic extinction effect has been removed from the \ebv\ values,
which thus strictly correspond to the intrinsic extinction within the source.
The mass of the central region SSC was obtained from the UV intrinsic
luminosity. Ratios between values for each region are also shown.}
\label{irasssctable}
\centering
\resizebox{\textwidth}{!}{
\begin{tabular}{lccccccc}
\hline\hline\\
Name & Region & \ebv$_{*}$ & \ebvneb\ & \fuviras\ & \lcuva$^{\mathrm{a}}$ &
\lha$^{\mathrm{a}}$ & $M$  \\
& & & & (\ergsacm) & (\ergsa) & (\ergs) & (\msun)  \\
\hline\\
STB & Integrated source & $0.15$ & $0.06$ & $4.5\times 10^{-14}$
& $3.0\times 10^{41}$ & $1.2\times 10^{42}$ & $1.4\times 10^{8}$ \\
SSC & Central stellar cluster & $0.01$ & $0.11$ & $8.5\times
10^{-15}$ & $1.5\times 10^{40}$ & $1.3\times 10^{41}$ & $7.0\times 10^{6}$ \\
\hline\\
Ratio (SSC/STB) & - & - & - & $20$\% & $5$\% & $10$\% & $5$\% \\
\hline
\end{tabular}
}
\begin{flushleft}
$^{\mathrm{a}}$ Luminosity value corrected for Galactic and intrinsic extinctions.
\end{flushleft}
\end{table*}

This disagreement might be due to the value we have assumed for the fraction $f$
of ionizing photons emitted by the massive stars of STB which are considered to
contribute to ionization. Hitherto we have assumed when using the predictions by
the CMHK02 models that $30$\% ($1-f$~$=0.3$) of ionizing photons are absorbed by
dust before they ionize any hydrogen atom. \citet{Mathis1971} and
\citet{Petrosian1972} obtained analytical expressions of $f$ as a function of
the optical depth $\tau$. Assuming the latter expression \citet{Degioia1992}
obtained values $1-f$~$\sim 0.3$ for regions in LMC, similar to those found by
\citet{Mezger1978} in the Galaxy. According to these results CMHK02 models
assume as an average value $(1-f)=0.3$ when calculating the intensities of the
nebular lines from the predicted number of ionizing photons produced by the
massive stars of the burst. However, values measured for $\tau$ have
uncertainties and might have a spread, which translates into a spread in the
parameter $1-f$. Actually, when assuming $1-f=0.5$, CMHK02 models predict a
value for \hal\ luminosity in IRAS~0833 in agreement with the value reported by
\citet{LopezSanchez06}. This value of $1-f$ is realistic, indeed similar to the
fraction measured in several Galactic H~II regions \citep{Mezger1978}, and
consistent with the rather high value of extinction measured in the STB
spectrum.

Infrared emission $FIR(40-120\; \mu \rm{m})$ was calculated using the relation
by  \citet{Helou85} based on the IRAS fluxes $F_{60\; \mu \rm{m}}=5.81$ Jy and
$F_{100\; \mu \rm{m}}=6.48$ Jy. Following \citet{Calzetti00}, the value of
$FIR(40-120\; \mu \rm{m})$ was converted to the estimated total far infrared
emission with the relation $FIR(1-1000\; \mu \rm{m})/FIR(40-120\; \mu \rm{m})
\sim 1.75$. The resulting value is \lfir~$=3.6\times10^{44}$ \ergs. Assuming
\ebv~$=0.15$ with the LMC extinction law and $1-f=0.5$, CMHK02 models predict
\lfir~$=4.2\times10^{44}$ \ergs\ for STB . Though the predicted \lfir\ value is
higher by 16\% than the observed one, both values are consistent considering the
inherent uncertainties of the method, and the systematic $\pm 10\%$ uncertainty
in the predicted luminosities discussed above. All the predicted and observed
values are shown in Table~\ref{irasinttable}.

The rather good agreement between the predicted and observed values of \lha\ and \lfir,
with a  general trend to some overestimation by the models, indicates that the UV
continuum (used for normalization) is contributed by the complete population of young
stars in the central region of IRAS~0833. Therefore, there are no evidences of additional
stellar knots potentially hidden within dense dust clouds, which would be invisible in the
UV, but would contribute to \lha\ and \lfir,  as found in other similar galaxies like
Haro~2 \citep{Oti12}.

\subsection{Underlying stellar population}

\label{irasunder}

\citet{Leitherer13} considered that the HST/COS aperture was large enough  ({\em
radius}$=1.25\arcsec$) to contain short-lived massive bursts at different ages,
mimicking a quasi-continuous star formation rate, and adopted a characteristic
age of 20~Myr. The detection by  \citet{Pellerin99} of the Ca~II
$\lambda\lambda8498$,~$8542$,~$8662$ \AA\ and Mg$_{2}$ $\lambda5177$ \AA\
absorption lines in an IR-optical spectrum of the source, which are signatures
of red (super)giants (RSGs) and late-type stars, respectively, supported in
principle this assumption, since  the presence of both O stars and RSGs requires
starbursts extended over at least $\sim$$10$ Myr. Nevertheless,
\citet{Pellerin99} attributed the strength of the Ca~II triplet and Mg$_{2}$
absorption lines to an old stellar population of $\ge 1$ Gyr, rather than to
young RSGs. As discussed by \citet{LopezSanchez06} it is known that IRAS~0833
has experienced previous star-formation episodes, with at least two older
stellar populations with ages $100-200$ Myr and $1-2$ Gyr, respectively.
\citet{MasHesse03} showed in their Fig.~11 that the optical continuum peaks
$\sim 1.2\arcsec$ ($\sim 470$ pc) north from a high-intensity emitting ionized
gas within the STIS slit. This region (which they labelled ``N'') presents a
strong Balmer discontinuity and a lack of ionizing power, and is consequently
dominated by older stars. We argue that, as we discuss below, the stars in this
region and other similar ones are dominating the features detected by
\citet{Pellerin99}, but do not have any relation with the young starburst in the
core of IRAS~0833.

Moreover, when we tried to reproduce the different observables assuming
synthesis models for extended bursts with the parameters considered by
\citet{Leitherer13}, the Si~ IV and C~IV spectral profiles could not be
simultaneously fitted, and the stellar mass value which would account for the UV
emission from STB overestimated the predicted \lha\ and \lfir\ by factors $7$
and $3$, respectively. As discussed above, the measured $WR/(WR+O)$ ratio is
neither consistent with an extended starburst. Therefore, our results reject the
hypothesis of a star formation episode extended over several tens of Myr in
IRAS~0833. We conclude that the Balmer lines emission, the UV continuum and the
far infrared luminosity are dominated by a starburst which converted $\sim
1.4\times 10^{8}$ \msun\ of gas into stars $\sim 5.5$ Myr ago. As already
explained, our data do not allow us to constrain the age of the young, massive
stars contributing to the UV continuum outside the central bar SSC, but the good
fit with the lower resolution global IUE spectrum indicates that the spread in
ages has to be very small. In addition, an underlying, older stellar population
which dominates the optical to infrared continuum must have been formed in
previous star formation episodes distributed over the last 1~Gyr.

In order to check this assumption, we plotted in Fig.~\ref{figirassed} (right
panel) the UV-optical SED of IRAS~0833. It includes the IUE spectrum already
fitted in the left panel and discussed in the text, as well as the photometric
fluxes obtained by \citet{Ostlin09} from HST/ACS observations. We noticed that
the same model described in Table~\ref{irasinttable} and which can reproduce the
UV slope and the UV, \hal\ and FIR emissions from STB, was unable to
account for the SED at $\lambda \gtrsim 4000$ \AA. Since there are evidences of
the presence of an older stellar population, as discussed above, we included a
1-Gyr-old SB99 model with same metallicity and mass limits as for the young
starburst. This spectrum was reddened by the same amount than the IUE continuum
to account for internal (\ebv = 0.15, LMC law) and Galactic (\ebv = 0.094,
Cardelli law) extinctions. As shown in Fig.~\ref{figirassed}, whereas the young
starburst population dominates the UV emission, the composite of the old and
young populations reproduces properly the optical fluxes reported by
\citet{Ostlin09} when $M_{\rm{old}}=6 \times 10^{10}$ \msun.

We have considered only a test case assuming a model for an age of $1$~Gyr. The
results do not vary significantly for other configurations, so that  we can
conclude that  a stellar population some hundreds of Myrs old contributes
significantly to the optical--IR continuum observed in IRAS~0833. Indeed,
Fig.~\ref{figirassed} shows that the continuum at $\lambda \sim 8500$ \AA\ is
dominated by the older stars, which therefore must account for the Ca~II triplet
absorption lines, as suggested by \citet{Pellerin99}. Finally, when considering
the contribution by the old stellar population, CMHK02 models predict
\ewhb~$\sim 14$ \AA\ for the integrated starburst in IRAS~0833. Bearing in mind
the uncertainties in the actual contribution of the underlying stellar
population to the stellar continuum, as well as the effect of extinction in its
optical emission, this \ewhb\ estimate is in agreement with the value
\ewhb~$=19$ \AA\ reported by \citet{LopezSanchez06}. The \ewhb\ calculated from
the models, as well as the observed value, are included in
Table~\ref{irasinttable}.

In brief, the UV, \hal\ and FIR emissions in IRAS~0833 are dominated by the
young stars produced in a starburst which occurred $\sim 5.5$ Myr ago, which
took place all along the central region of the galaxy, as well as in the
surrounding spiral arms. On the other hand, the optical--IR continuum, as well
as other features such as the Ca~II triplet, are mostly contributed by old stars
formed some hundred millions years before the current starburst took place.

\begin{figure*}
\begin{center}
\includegraphics[width=11 cm,bb=14 14 343 285
dpi,clip=true]{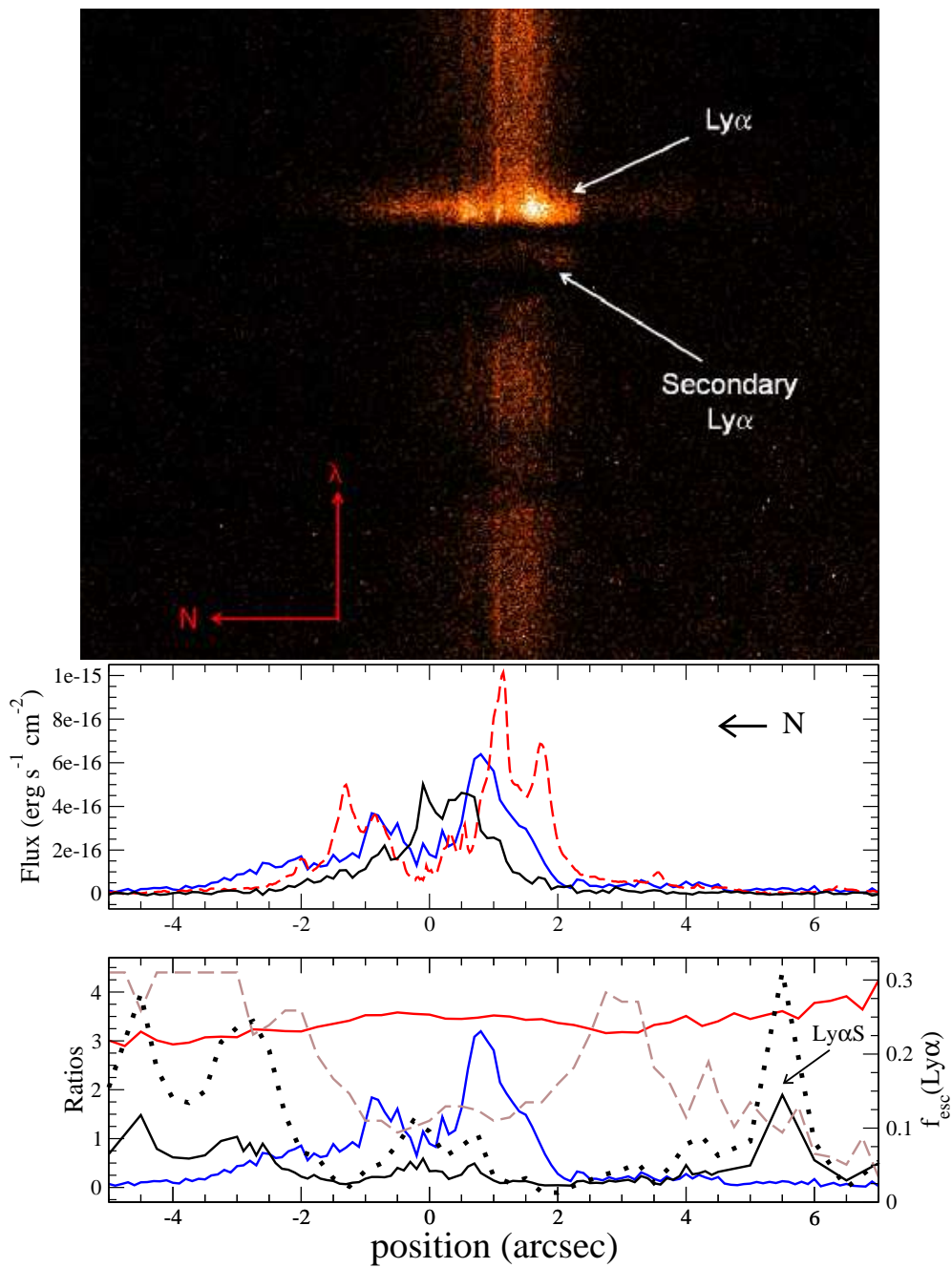}
\\
\includegraphics[width=13 cm,bb=12 39 758 522
dpi,clip=true]{fesc_Lya.05.print.eps}
\end{center}
\caption[Spectral images and emissions spatial profiles of IRAS~0833]{{\bf Top}:
HST/STIS G140M spectral image of IRAS~0833. Spatial extension is $\sim
19\arcsec$ ($\sim$$7.5$ kpc) . {\bf Middle}: emission profile of the main
component of \lyalp\ (blue) and UV continuum ($1224.5$ \AA\ rest frame, black)
from STIS data, and \hal\ (red-dashed) from ACS data. {\bf Bottom}: spatial
profiles of \lyalp\ (blue) and observed \lyalp$/$\hal\ (solid black) and
\hal$/$\hb\ (from WHT/ISIS data, in red), and the expected \lyalp$/$\hal\ ratio
(brown-dashed), assuming Case B recombination and internal reddening as derived
from the \hal$/$\hb\ ratio, as well as the Galactic extinction. The \lyalp\
escape fraction \fesclya, as defined in the text, has been plotted in
black-dotted line. In the middle and bottom panels the negative $x$-axis
corresponds to north, as indicated by {\em N}. To ease the comparison, the red
dashed line in the middle panel shows the \hal\ profile scaled by a factor
$0.2$. The vertical scale corresponds to the \lyalp\ and non-scaled \hal\
fluxes. Whereas the vertical scale of the left axis in the bottom panel
corresponds to the \lyalp$/$\hal\ and \hal$/$\hb\ ratios, the scale of the right
axis corresponds to \fesclya. Position of the region \lyas\ is marked.}
\label{figirasratios}
\end{figure*}

\subsection{\lyalp\ emission}

\label{iraslyares}

A high-resolution spectrum around \lyalp\ of the central region of IRAS~0833 is
shown in Fig.~\ref{irlya}. This 1-d spectrum was extracted by \citet{MasHesse03}
from the HST/STIS G140M spectral image, which is shown in
Fig.~\ref{figirasratios} (top panel). The \lyalp\ emission is very prominent,
extending well beyond the stellar continuum to the north direction, whereas the
emission toward the south is much weaker. Unlike what \citet{Oti12} found on the
major axis of Haro~2, regions of total \lyalp\ absorption are not observed along
the slit. The \lyalp\ spatial profile in Fig.~\ref{figirasratios} (middle panel)
shows two compact components at both sides of the central cluster: north
peak ({\em position}~$\sim -0.7\arcsec \sim -300$ pc, {\em width}~$\sim 1\arcsec
\sim 400$ pc, $F \sim 4\times10^{-16}$ \ergscm) , and south peak ({\em
position}~$\sim 1\arcsec \sim 400$ pc, {\em width}~$\sim 1.5\arcsec \sim 600$
pc, $F \sim 6\times10^{-16}$ \ergscm). In the region between them there is an
abrupt decline, which is co-spatial to the stellar cluster. Also, a somewhat
diffuse emission is observed extending to the north ({\em position}~$<-1\arcsec
\sim 400$ pc) over $\sim$$6\arcsec$ ($\sim$$2.3$ kpc), and with a
lower flux ($F < 2\times10^{-16}$ \ergscm). On the other hand, the emission
extending south from the southern peak is much weaker. The total spatial
extension of the \lyalp\ emission is $\sim$$16\arcsec$ ($\sim$$6$ kpc), whereas
the stellar continuum, dominated by the central stellar cluster, extends over
$\sim$$5\arcsec$ ($\sim$$1.9$ kpc). The observed, integrated flux of \lyalp\
along the STIS slit is $F($\lyalp$) \sim 6.4\times10^{-14}$ \ergscm\ (not
corrected for any kind of absorption or extinction), with the diffuse emission
contributing up to $\sim 25$\% of the total one.

\begin{figure}
\centering
\includegraphics[width=8 cm,bb=2 40 706 522
dpi,clip=true]{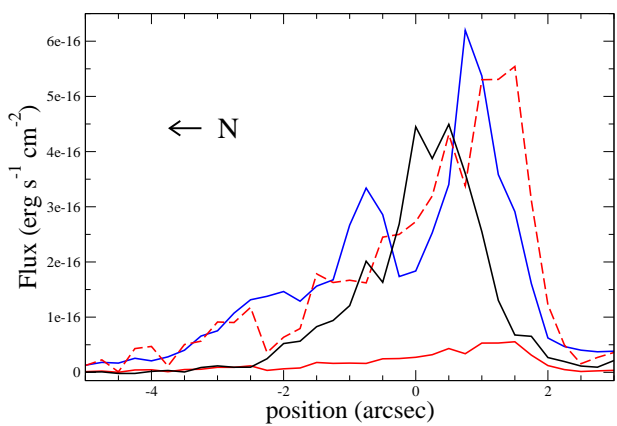}
\caption[Spatial profile of the secondary \lyalp\ emission of
IRAS~0833]{Emission profiles of the UV continuum ($1224.5$ \AA\ rest frame,
black), primary \lyalp\ (blue) and secondary \lyalp\ (red) emissions from
HST/STIS data. The vertical scale corresponds to the \lyalp\ flux. The secondary
emission has been scaled by a factor $10$ (red dashed line) to ease the
comparison. Spatial resolution of all the profiles was degraded to $0.25$
arcsec pixel$^{-1}$.}
\label{figirasseclya}
\end{figure}

\begin{figure}
\begin{center}
\includegraphics[width=8.5 cm,bb=42 41 755 508 dpi,clip=true]{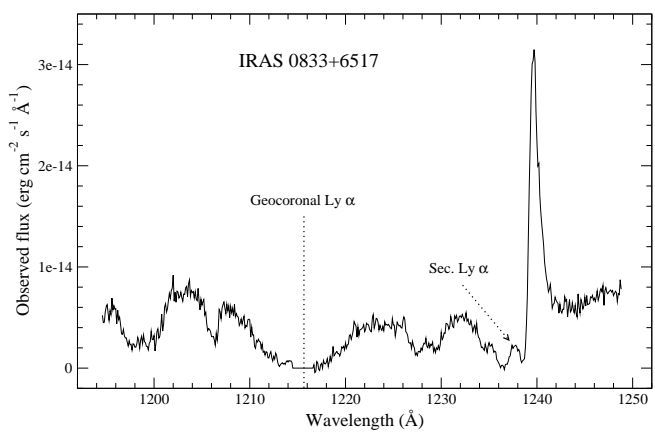}
\end{center}
\caption{Extracted 1-d spectrum of the central region of IRAS~0833
(corresponding to 2\farcs0). The position of the geocoronal \lyalp\ line has
been marked, in the center of the broad Galactic absorption profile. We have
also marked the position of the secondary emission peak discussed in the text.
Reproduced from \citet{MasHesse03}.}
\label{irlya}
\end{figure}

Figures~\ref{figirasratios} and \ref{irlya} show also a deep, broad \lyalp\
absorption bluewards of the emission. \citet{MasHesse03} concluded that this
P~Cyg profile was consistent with scattering by a neutral medium outflowing at
$\sim$$300$ km s$^{-1}$. A weak secondary \lyalp\ emission component is observed
onto the deep absorption trough. This emission is also blueshifted by
$\sim$$300$ km s$^{-1}$. Moreover, this secondary emission extends along
$\sim$$6.0\arcsec$ ($\sim$$2.3$ kpc), with a total observed flux of $F($\lyalp$)
\sim 5.5\times10^{-15}$ \ergscm. Figure~\ref{figirasseclya} shows the spatial
profile of the main and secondary \lyalp\ components, together with the profile
of the UV continuum. As observed, although $\sim$$10$ times weaker, the
secondary \lyalp\ emission roughly follows the spatial profile of the primary
component, except for the peak $\sim$$1\arcsec$ north of the central cluster,
which is not observed in the former. Both emissions are spatially decoupled from
the stellar continuum.

\citet{MasHesse03} found the same velocity structure in the P~Cyg profile
of the \lyalp\ emission detected in IRAS~0833 all along the STIS slit, i.e. over
at least $10\arcsec$ ($\sim$$4$ kpc). They found that the profiles fall to zero
at the same velocities within a narrow range ($\pm50$ km s$^{-1}$), regardless
of the distance to the central region. Therefore, these observations indicate
that most of the neutral gas responsible for the absorption feature must be
approaching us at the same velocity $\sim$$300$ km s$^{-1}$ all along the slit.

\begin{table*}
\caption[X-ray spectral fitting of IRAS~0833]{X-ray spectral fitting of
IRAS~0833, with the number of degrees of freedom ($\nu$) and the reduced
$\chi^{2}$ value ($\chi^{2}/\nu$).}
\label{xrayirasfit}
\centering
\begin{tabular}{lccccc}
\hline\hline\\
Model & \multicolumn{1}{c}{$kT$ (keV)} & \multicolumn{1}{c}{$kT$ (keV) /
$\Gamma$} & Metallicity & $\nu$ & $\chi^{2}/\nu$
 \vspace*{1 mm} \\
& Norm$^{\mathrm{a}}$ & Norm$^{\mathrm{a}}$ & $Z_\odot$ & & \\
\hline\\
1: wabs*zwabs*(zmekal+zmekal) & \multicolumn{1}{c}{$0.60_{-0.11}^{+0.06}$} &
\multicolumn{1}{c}{$4.2_{-1.0}^{+1.7}$} & $0.4$ & $37$ &
$1.056$ \vspace*{1 mm} \\
& $6.0_{-1.1}^{+1.1}\times10^{-5}$ & $1.44_{-0.18}^{+0.18}\times10^{-4}$ & & &
\vspace*{1 mm} \\
2: wabs*zwabs*(zmekal+zpowerlw) & \multicolumn{1}{c}{$0.62_{-0.11}^{+0.08}$} &
\multicolumn{1}{c}{$2.0_{-0.2}^{+0.2}$} & $0.4$ & $37$ &
$0.993$ \vspace*{1 mm} \\
& $4.6_{-1.5}^{+1.4}\times10^{-5}$ & $4.9_{-0.9}^{+1.0}\times10^{-5}$ & & &
\vspace*{1 mm} \\
\hline
\end{tabular}
\begin{flushleft}
$^{\mathrm{a}}$ Units of normalization as in XSPEC. Hot plasma: $10^{-14}/\{4\pi
[D_{A} (1+z)]^{2}\} \int n_{e} n_{H} \rm{d} V$. Power-law: photons s$^{-1}$
cm$^{-2}$ keV$^{-1}$ ($1$ keV).
\end{flushleft}
\end{table*}

\begin{table*}
\caption[X-ray fluxes and luminosities of IRAS~0833]{X-ray fluxes and
luminosities of IRAS~0833, together with the fractional contribution to the soft
X-ray range by the harder spectral components.}
\label{irasxraylum}
\centering
\resizebox{\textwidth}{!}{
\begin{tabular}{lccccc}
\hline\hline\\
Model & $F_{0.4-2.4\rm{\,keV}}$ $^{\mathrm{a}}$ & $F_{2.0-10.0\rm{\,keV}}$
$^{\mathrm{a}}$ & $L_{0.4-2.4\rm{\,keV}}$ $^{\mathrm{b}}$ &
$L_{2.0-10.0\rm{\,keV}}$ $^{\mathrm{b}}$ & \lxratt\ $^{\mathrm{b}}$ \vspace*{1
mm} \\
& (\ergscm) & (\ergscm) & (\ergs) & (\ergs) & \\
\hline\\
1: wabs*zwabs*(zmekal+zmekal) & $1.49_{-0.09}^{+0.08}\times10^{-13}$ &
$1.13_{-0.3}^{+0.18}\times10^{-13}$ & $1.36_{-0.08}^{+0.08}\times10^{41}$ &
$8.7_{-2}^{+1.4}\times10^{40}$ & $0.60_{-0.07}^{+0.07}$ \vspace*{1 mm} \\  
2: wabs*zwabs*(zmekal+zpowerlw) & $1.58_{-0.11}^{+0.11}\times10^{-13}$ &
$1.2_{-0.2}^{+0.2}\times10^{-13}$ & $1.47_{-0.10}^{+0.10}\times10^{41}$ &
$9.3_{-1.5}^{+1.7}\times10^{40}$ & $0.71_{-0.09}^{+0.11}$ \vspace*{1 mm} \\
\hline
\end{tabular}
}
\begin{flushleft}
$^{\mathrm{a}}$ Values of fluxes have not been corrected for neutral
absorption.\\
$^{\mathrm{b}}$ Values of luminosities have been corrected for neutral
absorption.
\end{flushleft}
\end{table*}

\subsection{X-rays}

As explained above, the analysis of the X-ray data of IRAS~0833 was performed
using {\em XSPEC}. The complexity of each model was increased once its
statistical significance had been checked with respect to the previous more
simple one. Changes in the models were considered valid if the probability
significance yielded by the F-test was $>99$\%. Galactic neutral absorption was
fixed to $N($H~I$)_{Gal}=4.5\times 10^{20}$ cm$^{-2}$ (see
Table~\ref{irastable}). The intrinsic neutral hydrogen column density was
measured by \citet{Kunth98} from the \lyalp\ profile, obtaining
$N($H~I$)_{int}=7.9\times 10^{19}$ cm$^{-2}$, although the uncertainties were
large and this value only corresponds to the line of sight of the central
cluster. We studied models with the parameter $N($H~I$)_{int}$ free, but no
improvement on the fitting was obtained over the models with $N($H~I$)_{int}$
fixed to the value reported by \citet{Kunth98}, which was therefore assumed.
This is due to the low intrinsic column density of IRAS~0833, which is roughly
one order of magnitude lower than the Galactic one. Metallicity was fixed to
$Z=0.4 Z_\odot$ as measured by \citet{LopezSanchez06}. As explained in
Sect.~\ref{irasobsxrays}, models were checked to yield similar values for the
parameters for both MOS1 and MOS2 spectra.

Models with only one emitting component did not yield satisfactory fittings.
Finally, two models were found for which the fits obtained were acceptable, and
which we labelled as Model~1 and Model~2. The former is a two-temperature hot
plasma (HP$_{\rm{soft}}$ and HP$_{\rm{hard}}$), modeled by {\tt mekal} in XSPEC
\citep{Mewe85}. On the other hand, Model~2 is a composite of a hot plasma and a
power-law emission (PL). All components from both models are affected by the
Galactic and the intrinsic absorptions. The values of the parameters obtained
for each of the models are listed in Table~\ref{xrayirasfit}, where errors were
calculated for a confidence level of $90$\%. The temperature of HP$_{\rm{soft}}$
is similar for Model~1 and Model~2 ($0.60_{-0.11}^{+0.06}$ keV and
$0.62_{-0.11}^{+0.08}$ keV, respectively), as well as its flux contribution,
which indicates that HP$_{\rm{soft}}$ accounts for the same emission in both
models. Only the modeling of the hard emission varies between them, being
another hot plasma (HP$_{\rm{hard}}$) with $kT=4.2_{-1.0}^{+1.7}$ keV in Model~1
and a power-law (PL) with $\Gamma=2.0_{-0.2}^{+0.2}$ in Model~2. When
considering a Raymond-Smith model ({\tt raymond} model in XSPEC
\citep{Raymond77}) instead of {\tt mekal}, a worse fitting was obtained, and
thus {\tt raymond} was rejected.

\begin{figure*}
\centering
\includegraphics[width=7.25 cm,bb=50 33 735 520
dpi,clip=true]{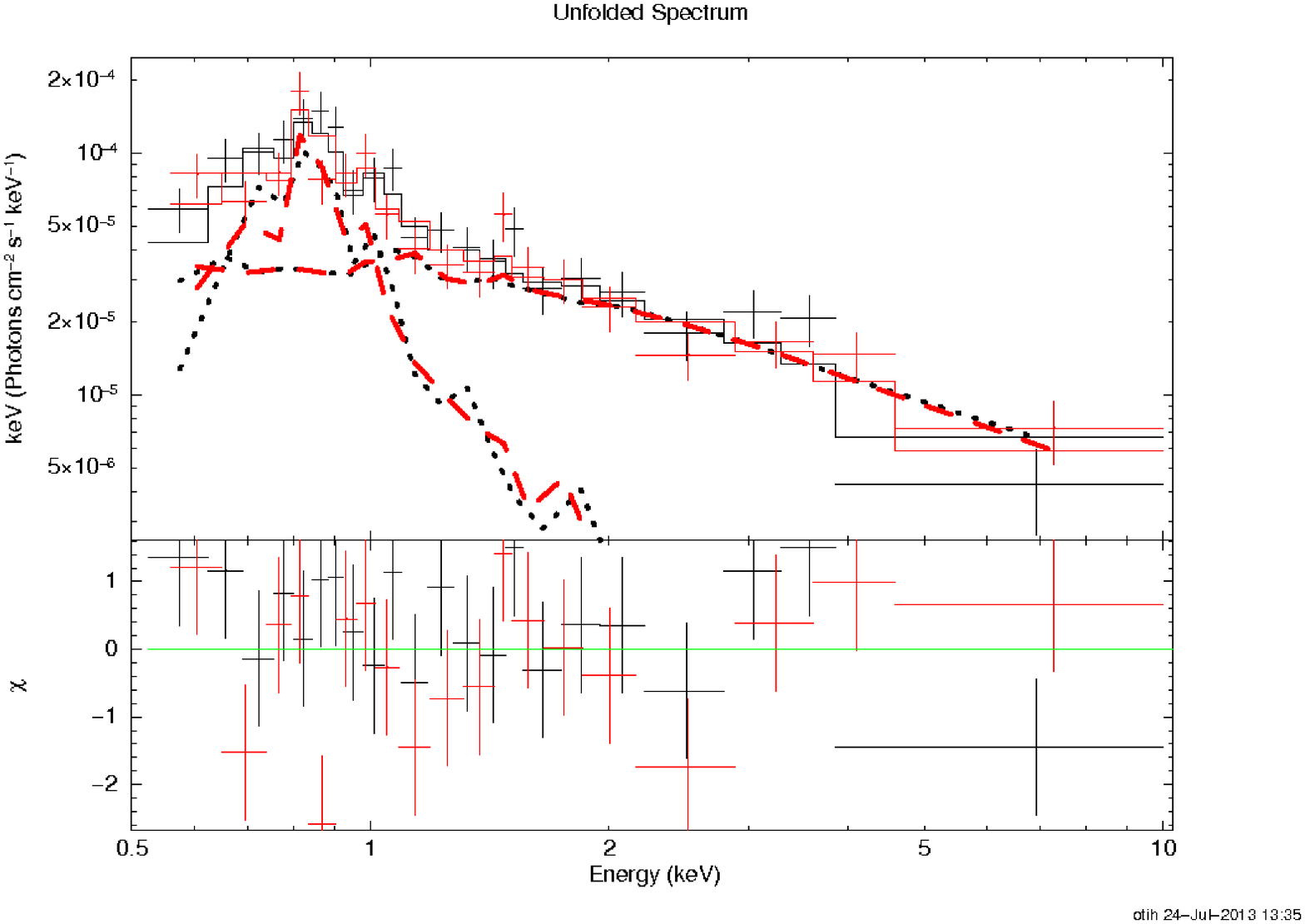}
\hspace{0.75 cm}
\includegraphics[width=7.25 cm,bb=50 33 735 520
dpi,clip=true]{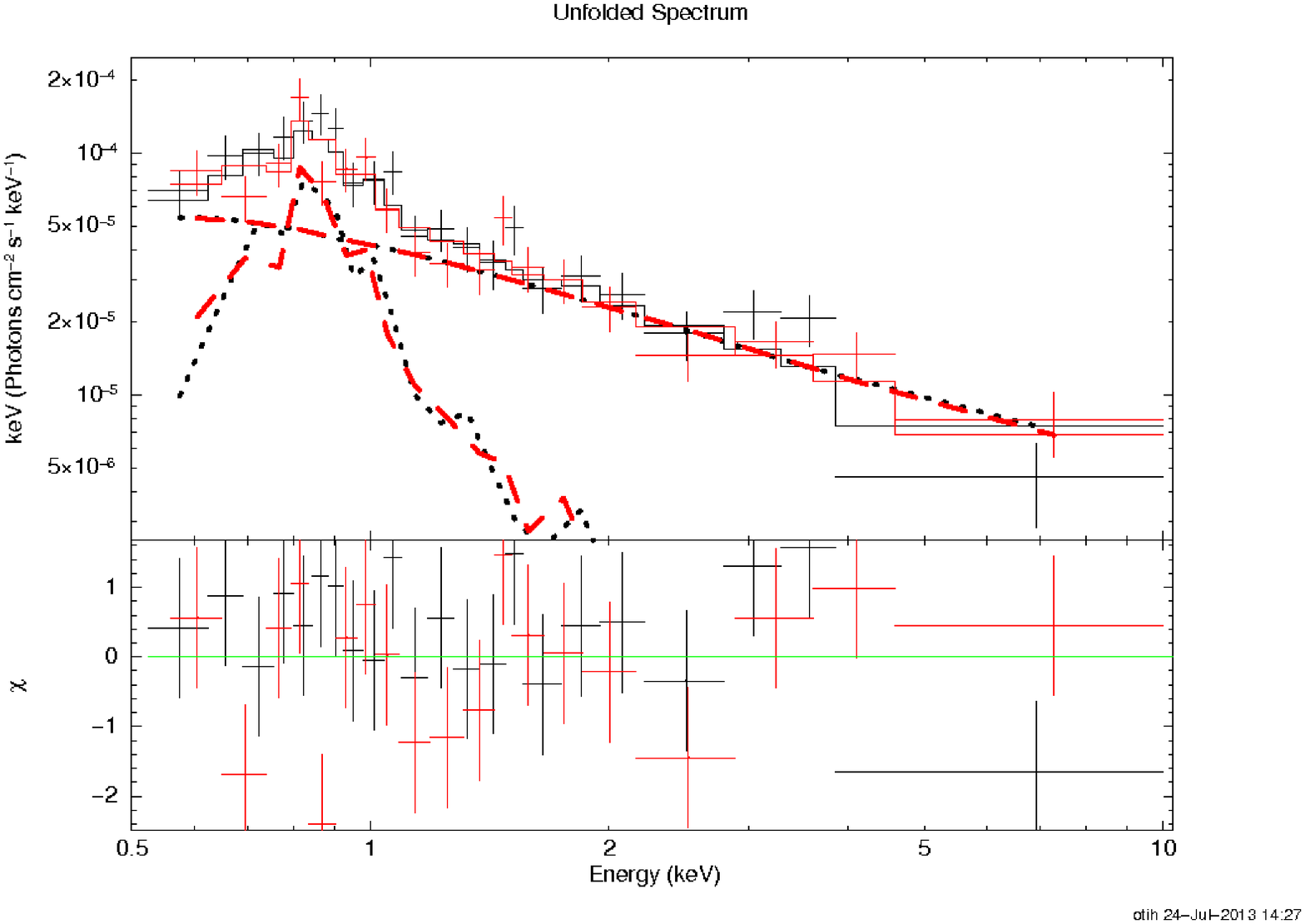}
\caption[X-ray spectrum of IRAS~0833]{XMM-Newton/EPIC X-ray spectrum of
IRAS~0833 with fitting Model~1 (left) and Model~2 (right). The different
components of the models are shown in dashed and dotted lines. MOS1 and MOS2
data are shown, together with the residuals of the fits (lower panels). Model~2
is favoured over Model~1 since it reproduces the hard X-ray emission with a
non-thermal component, as predicted by synthesis models}.
\label{figirasxspec}
\end{figure*}

In Table~\ref{irasxraylum} the values of the fluxes and luminosities for
Models~1 and 2 are shown, together with \lxratt, the fractional contribution to
the total soft X-ray ($0.4-2.4$ keV) luminosity by the component dominating the
hard X-ray range, i.e.
$L_{0.4-2.4\rm{\,keV}}^{\rm{HP_{hard}}}/L_{0.4-2.4\rm{\,keV}}^{\rm{tot}}$ for
Model~1, and $L_{0.4-2.4\rm{\,keV}}^{\rm{PL}}/L_{0.4-2.4\rm{\,keV}}^{\rm{tot}}$
for Model~2. Errors correspond to a confidence level of $90$\%. MOS1 and MOS2
spectra of the source are shown in Fig.~\ref{figirasxspec}, together with the
fitting Models~1 and 2. \citet{Stevens98B} analyzed a ROSAT/PSPC observation and
fitted the resulted spectrum of the source with a model of an absorbed hot
plasma. Assuming a Galactic column density of $N($H~I$)_{Gal}=4.08\times
10^{20}$ cm$^{-2}$ they obtained $kT=0.58_{-0.08}^{+0.6}$ keV,
$N($H~I$)_{int}=1.4_{-1.1}^{+5}\times 10^{21}$ cm$^{-2}$, $Z=0.02_{-0.02}^{+0.2}
Z_\odot$, and a value for the soft X-ray luminosity
$L_{0.1-2.5\rm{\,keV}}=2.8_{-1.7}^{+9}\times10^{41}$ \ergs. Uncertainties are
very high since the statistics of the ROSAT spectrum contained only $9$ points.
Our XMM--Newton data correct downwards the luminosity measured by
\citet{Stevens98B}.

\section{Discussion} 

\label{discussioniras}

We have analyzed UV-optical observational data of the local, face-on spiral
galaxy IRAS~0833, which hosts an intense \lyalp-emitting starburst. We have
concluded that the onset of the present starburst occurred $\sim$$5.5$ Myr ago,
leading to the production of a total stellar mass of $1.4 \times 10^{8}$ \msun.
While the stellar UV continuum of the integrated source (STB) is highly reddened
by dust, the emission of the super stellar cluster in its core (SSC) barely
shows any dust extinction. This irregular distribution of dust causes SSC to
account for $5$\% of the STB mass, but up to $20$\% of the integrated observed
UV flux. IRAS~0833 shows a prominent \lyalp\ emission, with a complex spectral
profile featuring two emission peaks. The brightest one shows a clear P~Cyg
profile due to resonant scattering by neutral gas outflowing at $\sim$$300$ km
s$^{-1}$. The $\sim$$10$ times weaker secondary \lyalp\ emission is observed on
top of the deep saturated absorption trough caused by the neutral hydrogen in
the main component. This secondary \lyalp\ emission appears blueshifted by the
same velocity $\sim$$300$ km s$^{-1}$ as the outflowing neutral gas. Both
compact and diffuse \lyalp\ emissions were detected along the slit. Whereas the
former is observed in two regions located between the central massive stellar
cluster and the \hal-emitting structures, the latter was detected extending
$\sim$$6\arcsec$ ($\sim$$2.3$ kpc) northward from the northern \lyalp\ compact
component, with a much weaker diffuse emission extending to the south. Finally,
we have analyzed the X-ray spectrum of STB, which is contributed by thermal
emission in the soft X-ray range, and a non-thermal power law, or a hotter
thermal component, extending also to the hard X-rays. The low spatial resolution
of the X-ray image does not allow to extract any morphological information.

In this section we will discuss the properties of SSC and its surroundings, the
origin of the X-ray luminosity from STB, and the distribution of the \lyalp\
emission over the central area of IRAS~0833.

\subsection{Central super stellar cluster}

\label{irasdisccentral}

We have seen in Fig.~\ref{figirasuvha} that the central super stellar cluster
(SSC) in IRAS~0833 shows a conspicuous UV continuum, but a very low local \hal\
emission. \lyalp\ shows also a local minimum at the position of the central SSC,
as shown in Fig.~\ref{figirasratios}. To analyze the lack of \hal\ emission in
the very inner regions of the IRAS~0833 core, a smaller circular region with
{\em radius}~$=0.5\arcsec$ ($\sim$$200$ pc, hereinafter \ri) was
considered, which is marked on Fig.~\ref{figirasuvha}. We extracted the \hal\
and UV fluxes of \ri\ and corrected them for the nebular and stellar dust
extinction, respectively, as discussed in Sect.~\ref{irasext}.  For an
instantaneous starburst with the properties given in Table~\ref{irasinttable}, 
and a mass which would account for the UV emission within \ri\ ($M \sim
3.1\times10^{6}$ \msun), CMHK02 models predict a value of \lha\ $\sim1.5$ times
larger than indeed observed within the aperture, even when assuming $1-f$~$=0.5$
as discussed above. In brief, around $30$\% of the ionizing photons emitted by
the SSC and not absorbed by dust, are actually escaping from the region.

\begin{figure}
\centering
\includegraphics[width=8 cm]{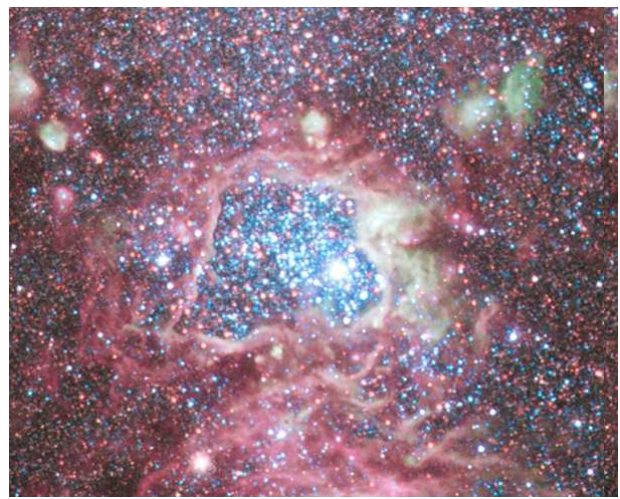}
\caption[N4214]{Central massive stellar cluster in the starburst galaxy NGC~4214, showing
the cavity formed around the massive stars. Adapted from the Hubble Heritage collection,
courtesy of R. O\textquoteright Connell and the WFC3 Scientific Oversight
Committee. }
\label{N4214}
\end{figure}

On the other hand, Fig.~\ref{figirasuvha} and the spatial profiles of the UV
continuum and \hal\ emission in Fig.~\ref{figirasratios} indicate that the
emitting nebular gas is mostly located towards the north and the south of the
central SSC. Specifically, we observe in Fig.~\ref{figirasratios} that, whereas
the bulk of the stellar cluster continuum is concentrated within $1\arcsec$
($400$ pc), \hal\ shows two main components: north peak ({\em position}~$\sim
-1.5\arcsec \sim -600$ pc, {\em width}~$\sim 1\arcsec \sim 400$ pc, $F \sim
2.5\times10^{-15}$ \ergscm), and south peak ({\em position}~$\sim 1\arcsec \sim
400$ pc, {\em width}~$\sim 1.5\arcsec \sim 600$ pc, $F \sim 5\times10^{-15}$
\ergscm). The weak \hal\ emission observed at the position of the SSC ($F \sim
2.5\times10^{-16}$ \ergscm) has to be due to the lack of gas within \ri. As
discussed above, the gas has been swept up by the intense starburst activity
(stellar winds, SNe, etc.) during the last $5.5$ Myr since the onset of the
burst, creating a cavity with a {\em diameter}~$\sim 1.5\arcsec$ ($\sim$$600$
pc). Outside this cavity \hal\ is actually prominent and structured, resulting
from ionization by the local massive stars and by the ionizing photons which
escaped from the central region. As already mentioned, the low extinction
derived from the stellar continuum in the core SSC indicates that the abundance
of dust within the cavity is also very low. These empty, dust--free cavities are
usual around strong starbursts, as discussed by \citet{Maiz1998} for NGC~4214.
Figure~\ref{N4214} shows in high detail how in these cases the nebular lines are
emitted by the gas swept up by the starburst, together with the dust, yielding a
strong differential extinction with respect to the stellar cluster, whose line
of sight remains almost free of dust and gas.

\subsection{X-rays}

\label{discxiras}

The low spatial resolution of the X-ray XMM-Newton/EPIC observations, with
IRAS~0833 $\sim 8 \arcmin$ away from the optical axis, did not allow to study
its morphology, but just to derive the integrated average physical properties of
the emitting gas, as discussed above.  The spectral analysis of IRAS~0833
unveiled the contribution by a hard component, which may be thermal (Model~1 in
Table~\ref{xrayirasfit}) or non-thermal (Model~2) in nature, and which in any
case contributes up to $\sim$$60-70$\% to $L_{0.4-2.4\rm{\,keV}}$. The rather
evolved state of the starburst in IRAS~0833 indicates that supernova remnants
(SNR) should already be present, and that a significant number of binary systems
might have already become active as high mass X-ray binaries (HMXBs), i.e. might
have started to accrete material onto a compact stellar remnant (black hole or
neutron star formed after the SN explosion of its massive, short-lived
progenitor). The CMHK02 models predict $N_{SN} \sim 10^{-9}$
\msun$^{-1}~\rm{yr}^{-1} $ \citep{Cervino94}, which scales to around 150 SN
explosions every $10^{3}$ yr for IRAS~0833, half of which are expected to be
part of a binary system. The fraction of them with also a massive secondary,
already in its last stages of evolution after 5.5 Myr, should be active as
HMXBs. The CMHK02 models predict $\sim 2 \times 10^{3}$ active HMXBs in
IRAS~0833 given the mass and evolutionary status of the starburst
\citep{Cervino97}. Assuming a typical luminosity of $L_{X} = 10^{38}$ \ergs\ for
a single HMXB, a contribution of $L_{X} = 2\times10^{41}$ \ergs\ by HMXBs is
expected. This value matches the luminosity contribution of the non-thermal
component of Model~2 of Table~\ref{xrayirasfit}. Furthermore, following the
method proposed by \citet{Mineo12a}, based on the calibration of HMXB luminosity
function in star--forming galaxies using a composite of \lfir\ and observed
\luv\ as a proxy, we find that $\sim60$ HMXBs with $L_{\rm{hardX}} > 10^{38}$
\ergs\ should be currently active in IRAS~0833, providing an integrated hard
X-ray luminosity $L_{0.5-8\rm{\,keV}} \sim 5 \times 10^{40}$ \ergs. While this
value is lower than the observed one, $L_{0.5-8\rm{\,keV}} \sim 2 \times
10^{41}$ \ergs, it is within $1 \sigma$ of the dispersion in the sample by
\citet{Mineo12a}. We conclude therefore that the hard X-ray component identified
in IRAS~0833 is consistent with the emission from the expected number of HMXBs
originated by the evolution of the starburst episode, favouring so the
non-thermal nature of the hard X-ray emission, as assumed in Model~2 of
Table~\ref{xrayirasfit}.

On the other hand, the soft X-ray component associated to the emission by a
thermal plasma at $kT \sim 0.62$ keV, as in Model~2, contributes with
$L_{0.4-2.4\rm{\,keV}} \sim 4.2 \times10^{40}$ \ergs. The soft X-ray emission in
starburst galaxies originates from 1) the heating of the diffuse gas surrounding
the stellar clusters up to temperatures of millions of Kelvin, due to the
stellar winds and supernovae injecting mechanical energy into the medium, and 2)
by the emission of supernova remnants (SNRs) during the adiabatic phase. The
contribution by the direct emission of individual stars is negligible when
compared to the previous components \citep{Cervino02b}. Evolutionary population
synthesis models CMHK02 compute the expected soft X-ray emission of a starburst
after defining the physical properties of the burst, considering that a free
fraction \xeff\ from the mechanical energy injected into the medium is finally
converted into soft X-ray luminosity. Typical values of \xeff$=0.01 - 0.1$
are found for star-forming objects \citep{Summers04,MasHesse08,Mineo12b}.

The CMHK02 models for a starburst episode as defined for IRAS~0833 in
Table~\ref{irasinttable} could reproduce the total observed soft X-ray
luminosity $L_{0.4-2.4\rm{\,keV}} \sim 1.47 \times10^{41}$ \ergs\ assuming
\xeff~$=0.03-0.04$. If we remove the contribution to the soft X-ray luminosity
by the non-thermal power law component,  then the efficiency would be
constrained to \xeff~$=0.01-0.02$. Nevertheless, since the extrapolation of the
hard X-ray component to the soft X-ray range might be quite uncertain, we
conclude that between $1\% - 4\%$ of the mechanical energy released by the
starburst in IRAS~0833 is heating the surrounding medium, being reemitted in the
form of soft X-ray photons. This value is within the usual range in starburst
galaxies, as commented above.  Indeed, the \lsxf\ ratio measured in IRAS~0833,
$log(L_{\rm{soft X}}/L_{\rm{FIR}}) = -3.9$, is very close to the mean ratio
measured by \citet{MasHesse08} on their complete sample of star--forming
galaxies, $log(L_{\rm{soft X}}/L_{\rm{FIR}}) = -3.92$ (see Fig.~5 in
\citet{MasHesse08}). We conclude therefore that the integrated soft and hard
X-ray emissions of IRAS~0833 are consistent with the predictions for a starburst
with the properties we have derived, so that no additional sources of X-rays
(e.g., a low-luminosity active galactic nucleus) have to be claimed to explain
them.

\subsection{\lyalp\ emission}

\label{disclyairas}

We show in Fig.~\ref{figirasratios} the spatial profiles of the UV continuum,
\lyalp\ and \hal\ emissions for IRAS~0833, together with the observed values of
the \hal/\hb\ and \lyalp/\hal\ ratios. The expected \lyalp/\hal\ has also been
included, calculated assuming Case B recombination (\lyalp/\hal~$=8.7$,
\hal/\hb~$=2.87$ \citep{Dopita03}) and applying the dust extinction derived
from the Balmer decrement. In what follows the effect of the Galactic extinction
was removed before the analysis. The values of color excess, line ratios or
escape fractions do not include this effect. The profiles were binned spatially,
lowering the spatial resolution when needed, in order to get always a
signal-to-noise ratio larger than $3$. Resulting resolutions are within the
range  $0.1 - 0.5$ arcsec pixel$^{-1}$.  We have also plotted in the bottom
panel of this figure the profile of the \lyalp\ escape fraction, as defined by
\citet{Atek08}, \\

$f_{\mathrm{esc}}^{\mathrm{Ly}\alpha}=L_{\mathrm{Ly}\alpha}^{obs}/L_{\mathrm{Ly}\alpha}^{
int}= 
L_{\mathrm{Ly}\alpha}^{obs}/(8.7\times L_{\mathrm{H}\alpha}^{obs}\times10^{0.4\cdotp
E_{B-V}\cdotp k_{6563}})$ \\

Therefore, $1-$\fesclya\ represents the fraction of \lyalp\ photons which are
destroyed internally (or at least scattered out of the line of sight), either by
dust absorption or by multiple interaction with the neutral gas. As we discuss
below, resonant scattering redistributes the \lyalp\ photons over large areas,
so that local \fesclya\ values may differ significantly from globally integrated
ones.

\begin{figure}
\centering
\includegraphics[width=9 cm, bb=13 39 706 523
dpi,clip=true]{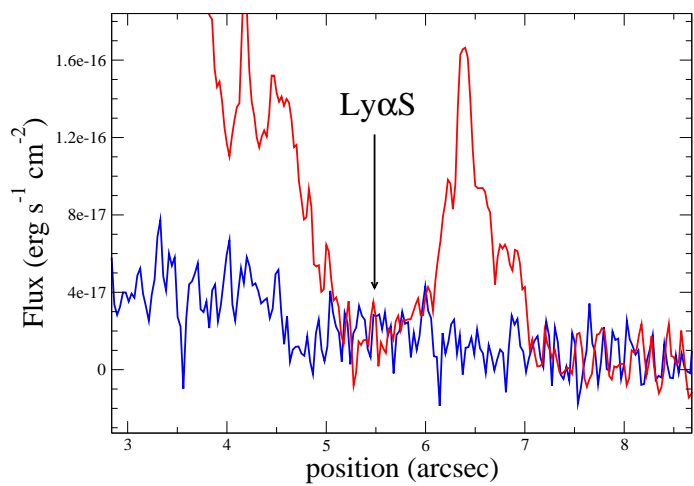}
\caption[P1]{Detail of the \lyalp\ (blue) and \hal\ (red) spatial profiles
around the region at $\sim5.5\arcsec$ ($\sim$$2.1$ kpc) south of the
central SSC showing the highest value of the \lyalp/\hal\ ratio (\lyas). The
profiles have been plotted at their original resolution, $0.029$ arcsec
pixel$^{-1}$ for \lyalp\ and $0.025$ arcsec pixel$^{-1}$ for \hal.}
\label{P1}
\end{figure}

The spatial profiles in Fig.~\ref{figirasratios} show that in addition to the
compact, central emitting blobs, a diffuse and extended  \lyalp\ component is
present along the STIS slit in IRAS~0833, as discussed in
Sect.~\ref{iraslyares}. Neither of them show any clear spatial correlation with
the UV continuum, Balmer decrement or \hal\ emission whatsoever, similarly as
found in Haro~2 \citep{Oti12}. As we concluded in Sect.~\ref{irasdisccentral},
the nebular gas seems to have been pushed out by the stellar activity, creating
a shell around the central, massive super-cluster SSC. This shell is being
ionized by SSC and is originating most of the \hal\ emission.
Figure~\ref{figirasratios} shows that the brightest \lyalp\ emitting regions are
also located around the central cluster. Nevertheless, there is a clear spatial
decoupling between the \lyalp\ and \hal\ profiles, as evidenced by the profile
of the \lyalp/\hal\ ratio and \fesclya\ over the nuclear region, with a local
maximum within the central $3\arcsec$ ($\sim$$1$ kpc). On the other hand, the
highest values of \lyalp/\hal\ ratio and \fesclya\ are detected in regions
dominated by diffuse emission northward  of the stellar cluster, at distances
higher than $2\arcsec$ ($\sim$$800$ pc), and in a very localized region at
$\sim$$5.5\arcsec$ ($\sim$$2.1$ kpc) to the south (\lyas, see
Figs.~\ref{figirasratios} and ~\ref{P1}). The location of \lyas\ at the
outskirts of IRAS~0833 nuclear region has been marked in
Fig.~\ref{figirasaper}.

\citet{Oti12} analyzed the \lyalp\ emission in Haro~2 and found also both
compact and diffuse components, neither of them spatially coupled to UV
continuum, Balmer emission or dust extinction as traced by \hal/\hb. However,
they found that 1) diffuse \lyalp\ emission was co-spatial to the diffuse soft
X-ray emission produced by the injection of mechanical energy into the ISM by
the starburst, and 2) the intensity of the diffuse \lyalp\ emission was higher
than the values predicted from \hal\ emission assuming Case B recombination and
correcting for dust extinction using the Balmer decrement. \citet{Oti12} argued
that this diffuse emission in Haro~2 might be originated in gas ionized by the
hot plasma responsible for the soft X-ray radiation, therefore not ionized by
the young massive stellar clusters in the central region. Recombination in this
environment would take place far from Case B conditions, causing 1) values of
\lyalp/\hal\ similar to those of Case B, but with 2) large values of \hal/\hb,
mimicking the presence of dust, and 3) values of the \lhalx\ ratio much lower
than when ionization is driven by massive stars. This model for the diffuse
\lyalp\ emission could explain the observed ratios in Haro~2.

But the scenario seems to be quite different in IRAS~0833. While the highest
values of the \lyalp/\hal\ ratios in IRAS~0833 occur also in regions dominated
by the diffuse emission, \lyalp\ emission marginally over the values expected
from Case B recombination is observed only at region \lyas. Unfortunately, no
X-ray image with enough spatial resolution is available to study 1) the
co-spatiality of the diffuse soft X-ray emission and the components of \lyalp,
and 2) the local values of the ratio \lhalx. Thus, we can not analyze in this
case whether the hot plasma could contribute significantly to the ionization of
the gas in some regions, as done for Haro~2. On the other hand,  the bulk of the
\lyalp\ emission comes from the region where \hal\ is also emitted, around the
central stellar cluster. The P~Cyg profile of the \lyalp\ line along the STIS
slit indicates that a large fraction of the \lyalp\ photons are being resonantly
scattered out of the line of sight by the outflowing neutral gas. Relatively
small variations in the properties of this outflow from one location to another,
might explain why the resulting \lyalp\ spatial profile differs from the \hal\
one in the central $3\arcsec$ area ($\sim$$1$ kpc). The scattering of a
significant fraction of the \lyalp\ photons explains indeed the low values of
the  \lyalp/\hal\ ratio along the STIS slit.

This scattering by the neutral gas surrounding SSC could transport the \lyalp\
photons to the outer regions of IRAS~0833, leaking from the galaxy as a weak,
diffuse emission halo. Fig.~\ref{figirasuvha} shows in its bottom right panel
the \lyalp\ image produced by \citet{Ostlin09}. While the procedure used to
extract the \lyalp\ emission in IRAS~0833 might be quite uncertain where the
stellar continuum is strong (and the \lyalp\ line shows a strong P~Cyg profile),
it provides a fair representation of the diffuse \lyalp\ emission. This diffuse
emission is quite homogeneous and does not follow the spatial structure of the
\hal\ emission.  A good example is found in region \lyas, as shown in
Fig.~\ref{P1}, where \lyalp/\hal\ becomes higher than 1 due to the local lack of
\hal\ photons. Eastwards of the nucleus, where the \hal\ emission is less
prominent, the diffuse \lyalp\ remains also at a similar surface brightness than
in other areas. The morphology of the \lyalp\ emission supports our conclusion
that the \lyalp\ photons which are missing along the slit are indeed smoothly
redistributed over the whole central area of IRAS~0833. Moreover, after refining
the method used by \citet{Ostlin09} to remove the UV continuum from the
$cont.+$\lyalp\ images, \citet{Hayes13b} obtained \lyalp\ images of the $14$
star-forming galaxies of the LARS sample \citep{Hayes13a}. Half of the sources
show a conspicuous \lyalp\ halo which do not present any spatial coupling to UV
or \hal\ emissions whatsoever. \citet{Hayes13b} argued that resonant scattering
may have redistributed the \lyalp\ photons over the large areas observed. A
similar situation seems to be present in IRAS~0833.

Combining the IUE \lyalp\ line flux and the value of the \hal\ integrated
emission measured by \citet{LopezSanchez06} we derived a globally integrated
ratio \lyalp/\hal$=0.4$ (corresponding to \fesclya\ $\sim 0.04$). This value is
$\sim 10$ times lower than the expected one assuming case B conditions and the
average nebular extinction, \lyalp/\hal$=4.4$. The destruction of \lyalp\
photons is therefore significantly larger than the extinction extrapolated from
the Balmer lines. Multiple resonant scattering of \lyalp\ photons by neutral
atoms increases significantly their probability of being destroyed by dust
grains, yielding a much higher extinction by dust for photons around 1216~\AA.
As predicted by \lyalp\ photons transfer models \citep{Verhamme06}, multiple
scattering can also end up shifting their energy, becoming diluted at the wings
of the emission line; this effect is negligible when dust is abundant, as in the
central area of IRAS~0833. This redistribution and spatial smoothing of the
\lyalp\ photons are responsible for the spatial decoupling between \lyalp\ and
\hal\ emissions visible in Figs.~\ref{figirasuvha} and \ref{figirasratios}. This
leads to the conclusion that the star formation rate may be severely
underestimated when based solely on integrated fluxes of the \lyalp\ line
without a proper correction of the aforementioned effects. See Fig.~2 in
\citet{Hayes11}, for instance.

In any case, we do not find evidences in IRAS~0833 of any further source of
\lyalp\ photons like ionization by hot plasma, as it was the case in Haro~2. If
present, its contribution should be negligible when compared to the ionization
by massive stars. Comparison with the results on Haro~2 brings us once again to
the evidence that the physics of \lyalp\ photons production, transfer and
destruction is extremely dependent on the conditions of each galaxy, being it
difficult to find any clear correlation with the macroscopic properties of each
object.

Based on the work by \citet{TenorioTagle99}, \citet{MasHesse03} proposed an
evolutionary picture of starbursts to try to explain the presence or lack of
\lyalp\ emission in star-forming galaxies, as well as the spectral features of
the line. They argued that the starburst activity in IRAS~0833 has accelerated
the surrounding material, forming a superbubble which expands at a velocity of
$\sim300$ km s$^{-1}$ in whose interior massive stars have ionized the gas.
Nebular emission lines are produced after recombination, \lyalp\ among them. Due
to its resonant nature, the bluest photons of the \lyalp\ line are scattered by
the neutral hydrogen on the shell of the superbubble, originating the P~Cyg
profile of the primary \lyalp\ component. Furthermore, they proposed that the
shock produced in the leading front may cause the shocked gas to undergo
recombination. Since no dense neutral gas layers are ahead of this front, the
\lyalp\ photons would not be scattered and would be detected by the observer
with a blueshift corresponding to the expansion velocity of the superbubble.
\citet{MasHesse03} proposed that the secondary \lyalp\ component in IRAS~0833
could be produced by this mechanism. This secondary emission does not seem to be
affected by neutral hydrogen scattering, which indicates that the H~I column
density in front of the expanding superbubble must indeed be low. This
blueshifted emission is in any case weaker by an order of magnitude than the
primary component, and does not contribute significantly to the integrated
\lyalp\ emission. According to the \lyalp\ spectral profiles observed, the
starburst in IRAS~0833 seems to be in the phase (d) of model by
\citet{TenorioTagle99}, as shown in their Fig.~8, which is characterized by a
\lyalp\ emission with a P~Cyg profile, together with a blueshifted emission
component.

On the other hand, \citet{Leitherer13} obtained an HST/COS spectrum of IRAS~0833 which
includes the \lyalp\ line. They fitted its spectral profile using the radiation transfer
code described by \citet{Schaerer11}, which is an improved version of the models developed
by \citet{Verhamme06}. Assuming the simple geometry of an expanding spherical shell, with
expansion velocity, Doppler parameter, column density and dust optical depth as free
parameters, \citet{Leitherer13} obtained a spectral profile which nicely reproduces the
primary \lyalp\ spectral component and predicted the presence of a secondary emission.
Nevertheless, the model fails to reproduce properly the wavelength of the secondary
emission peak, as well as the profile of the blue, P Cyg-like absorption wing. More
detailed modeling of the \lyalp\ line in IRAS~0833 would be needed to unveil the origin of
this secondary \lyalp\ emission.

\section{Conclusions}
\label{conclusions}

We have carried out a multiwavelength spectral and photometric analysis of the
face-on, spiral, star-forming galaxy IRAS~0833. The properties of its starburst
episode and associated X-ray emission have been defined, and the morphology of
the \lyalp\ emission has been discussed, together with its relation to the UV
continuum from the massive stars, the \hal\ line and the \hal/\hb\ ratio.

\begin{itemize}

\item The properties of the starburst in IRAS~0833 have been defined by
comparison to evolutionary population synthesis models. We conclude that a
single stellar population with an age $5.5$ Myr, a mass of $M \sim
1.4\times10^{8}$ \msun, and average stellar and nebular extinctions of
\ebv=$0.15$ and \ebv=$0.06$, respectively, can account for the integrated UV,
\hal\ and FIR photometry, as well as for the spectral profile of the
photospheric Si~IV and C~IV lines. The integrated UV spectrum shows the $2175$
\AA\ feature, consistent with an LMC-like extinction law.

\item An underlying stellar population, some hundreds of Myrs old, dominates the
IR-optical continuum in IRAS~0833, as well as the observed Ca~II triplet and
Mg$_{2}$ absorption lines.

\item The nucleus of IRAS~0833 shows a massive cluster whose stellar continuum
is barely attenuated (\ebv=$0.01$). We have estimated that it accounts for
$20$\% of the observed integrated UV flux and $5$\% of the total mass of the
starburst. The images of the nucleus show a very weak \hal\ emission co-spatial
to the central cluster. In fact, the synthesis models overestimate by a factor
$1.5$ the \hal\ luminosity emitted within a central circular region of {\em
radius}~$=0.5\arcsec$ ($\sim$$200$ pc). We argue that the low \hal\ emission
observed in this core region is due to the lack of nebular gas around the
massive central cluster, which has cleaned its surroundings after pushing out
the nearby gas, producing an expanding \hal-emitting shell.

\item The X-ray image does not provide any spatial information on the morphology of the
emission. The starburst episode can account for the thermal soft X-ray emission detected
assuming an efficiency in the conversion of mechanical energy into X-rays of
$\sim$$1-4$\%, within the typical range for starburst galaxies. The non-thermal component,
which dominates the hard X-ray emission, is consistent with the number of active HMXBs
expected for an evolved starburst with the properties derived for IRAS~0833.

\item Two spectral \lyalp\ components are observed. The main \lyalp\ is detected
along $16\arcsec$ ($\sim$$6$ kpc) and shows a P~Cyg profile originated by
resonant scattering by neutral hydrogen outflowing at $\sim$$300$ km s$^{-1}$.
The velocity structure extends all along the slit. The secondary emission is
$\sim$$10$ times weaker, extends over $6\arcsec$ ($\sim$$2.3$ kpc) and is
detected over the spectral trough caused by the outflowing neutral material. All
this indicates that the starburst activity has accelerated the sourrounding
neutral gas which no longer scatters the whole line, but only its bluest
photons. In the leading front of the superbubble the gas seems to have undergone
recombination, producing the secondary \lyalp\ emission, with probably scarce
neutral material ahead, so that it is detected blueshifted by $\sim$$300$ km
s$^{-1}$.

\item \lyalp\ is observed in emission along $16\arcsec$ ($\sim$$6$ kpc), without
regions of total \lyalp\ absorption. It does not show any spatial matching with
the UV continuum, \hal\ line or Balmer decrement whatsoever. \lyalp\ photons
escape mostly from two compact areas associated to, but not coincident with, the
\hal-emitting shell, with a size of $1\arcsec$ ($\sim$$400$ pc, north) and
$1.5\arcsec$ ($\sim$$600$ pc, south). In addition, there is a diffuse \lyalp\
component which extends northward $6\arcsec$ ($\sim$$2.3$ kpc).

\item  Similarly as found in other starburst galaxies, like Haro~2
\citep{Oti12}, the diffuse \lyalp\ emission shows the highest \lyalp/\hal\
ratios, as well as the highest values of the \lyalp\ escape fraction, \fesclya.
The globally integrated \lyalp/\hal\ ratio is well below the Case B predictions
(even after dereddening using the \ebvneb\ average value derived from the
\hal/\hb\ ratio), indicating an enhanced destruction of \lyalp\ photons with
respect to the extinction extrapolated from the Balmer lines.

\item We conclude that the outflowing neutral gas in front of the IRAS~0833
starburst area is resonantly scattering most of the \lyalp\ photons, producing
the observed P~Cyg profile. These scattered photons are smoothly redistributed
over the whole central area until either they are destroyed by interaction with
dust, or escape from an extended region. Multiple scattering enhances the
probability of interaction with dust, explaining why the \lyalp/\hal\ ratio
values are much lower than expected for Case B conditions and the average
extinction. We do not find any evidence in IRAS~0833 of any other further source
of \lyalp\ photons, like ionization by a hot plasma as proposed for Haro~2. If
present, its contribution should be negligible when compared to the ionization
by massive stars.

\item  The star formation rate in galaxies, e.g. high-redshift sources, may be
severly underestimated when it is derived only from integrated fluxes of the
\lyalp\ line without a proper correction of transfer effects in the interstellar
medium.

\end{itemize}

\begin{acknowledgements}

HOF and JMMH are partially funded by Spanish MINECO grants AYA2010-21887-C04-02
({\em ESTALLIDOS}), AYA2011-24780/ESP and AYA2012-39362-C02-01. HOF is funded by
Spanish FPI grant BES-2006-13489, CONACYT grant 129204 and a postdoctoral UNAM
grant. G.\"{O} is a Swedish Royal Academy of Sciences research fellow supported
by a grant from Knut and Alice Wallenberg foundation, and also acknowledges
support from the Swedish Research Council (VR) and the Swedish National Space
Board (SNSB). MH received support from the Agence Nationale de la Recherche
(ANR-09-BLAN-0234-01). HA and DK are supported by the Centre National d'\'Etudes
Spatiales (CNES) and the Programme National de Cosmologie et Galaxies (PNCG).
This research has made use of the Spanish Virtual Observatory
(http://svo.cab.inta-csic.es) supported from the Spanish MICINN / MINECO through
grants AyA2008-02156, AyA2011-24052. We want to acknowledge the use of the {\em
Starburst99} models and the NASA/IPAC Extragalactic Database (NED). We are very
grateful to the {\em ESTALLIDOS} collaboration for its scientific support. This
paper was based on observations with {\em Hubble Space Telescope}, {\em
XMM-Newton}, {\em International Ultraviolet Explorer} and {\em William Herschel
Telescope}. We thank the anonymous referee for his comments, which helped to
improve the understandability and clarity of the manuscript.

\end{acknowledgements}

\end{document}